\documentclass[a4paper,12pt]{article}
\usepackage[dvipdfmx]{graphicx}
\graphicspath{{./}} 
\usepackage{amsmath, amssymb}
\usepackage{color}
\usepackage{bm}
\usepackage[margin=2.4cm]{geometry} 
\usepackage{setspace} 

\usepackage{amssymb}
\usepackage{amsmath}
\usepackage{cancel}
\usepackage{ulem}

\title{Partial-wave decomposition on the lattice and its applications to the HAL QCD method}
\author{
	Takaya~Miyamoto${}^{1,2}$,
	Yutaro~Akahoshi${}^{1,2}$,
	Sinya~Aoki${}^{1,2}$, 
	Tatsumi~Aoyama${}^{2,3,4}$, \\
	Takumi~Doi${}^{2,5}$, 
	Shinya~Gongyo${}^{2}$, and
	Kenji~Sasaki${}^{1,2}$ \\
	{\small ${}^1$Center for Gravitational Physics, Yukawa Institute for Theoretical Physics,} \\
	{\small Kyoto University, Kyoto 606-8502, Japan} \\
	{\small ${}^2$RIKEN Nishina Center (RNC), Saitama 351-0198, Japan} \\
	{\small ${}^3$Institute of Particle and Nuclear Studies, High Energy Accelerator Research Organization(KEK),} \\
	{\small Tsukuba, Ibaraki 305-0801, Japan} \\
	{\small ${}^4$Kobayashi-Maskawa Institute for the Origin of Particles and the Universe (KMI),} \\
	{\small Nagoya University, Nagoya, 464-8602, Japan} \\
	{\small ${}^5$RIKEN Interdisciplinary Theoretical and Mathematical Sciences Program (iTHEMS),} \\
	{\small Saitama 351-0198, Japan}
}
\date{}
\begin{document}
\maketitle
\begin{abstract}
	The approximated partial wave decomposition method to the discrete data on a cubic lattice, developed by C.~W.~Misner, is applied to the calculation of $S$-wave hadron-hadron scatterings by the HAL QCD method in lattice QCD.
	We consider the Nambu-Bethe-Salpeter (NBS)  wave function for the spin-singlet $\Lambda_c N$ system calculated in the $(2+1)$-flavor QCD on a $(32a~\mathrm{fm})^3$ lattice at the lattice spacing $a\simeq0.0907$ fm and $m_\pi \simeq 700$ MeV. 
	We find that the $l=0$ component can be successfully extracted by Misner's method from the  NBS wave function projected to $A_1^+$ representation of the cubic group, which contains small $l\ge 4$ components.
	Furthermore, while  the higher partial wave components are enhanced so as to produce significant comb-like structures in the conventional HAL QCD potential if the Laplacian approximated by the usual second order difference is applied to the NBS wave function, such structures are found to be absent in the potential extracted by Misner's method, where the Laplacian can be evaluated analytically for each partial wave component.
	Despite the difference in the potentials, two methods give almost identical results on the central values and on the magnitude of statistical errors for the fits of the potentials, and consequently on the scattering phase shifts.
	This indicates not only that Misner's method works well in lattice QCD with the HAL QCD method but also that the contaminations from higher partial waves in the study of $S$-wave scatterings are well under control even in the conventional HAL QCD method. 
	It will be of interest to study interactions in higher partial wave channels in the HAL QCD method with Misner's decomposition, where the utility of this new technique may become clearer.
\end{abstract}
\section{Introduction} \label{sec:intro}
A determination of hadron-hadron interactions from the first-principle is one of the ultimate goals in both particle and nuclear physics. 
In a lattice QCD calculation of a two-hadron system,
the quantum numbers of the system are specified by the source and/or sink operators in the corresponding correlation function.
Among quantum numbers, the partial wave can be specified by the angular dependence in terms of the relative coordinate of two hadron operator, in principle. 
In lattice QCD, however, the extraction of the designated partial wave becomes non-trivial, because the rotational symmetry is broken to the cubic symmetry due to the finite volume (IR-effect) as well as the finite lattice spacing (UV-effect),
which introduces the mixing between different partial waves \cite{Luscher:NPB1991,Murano:PTP2011}.
In addition, full account of the angular dependence cannot be obtained since the data are available only on discretized spacial coordinates.

In Ref.~\cite{Misner:CQG2004}, a general method to (approximately) obtain a radial function in the particular partial wave on the cubic lattice was proposed by C.~W.~Misner, and it has been applied to the analyses of gravitational waves simulated on the grid points \cite{Fiske:RPD2005}.
Applying this method to lattice QCD enables us to 
verify how much the partial wave mixing is induced by the breaking of
the rotational symmetry and evaluate the corresponding systematics in the calculations of hadron-hadron interactions. 
In addition, this method could open a new possibility to extract interactions in higher partial waves
which are otherwise difficult to be studied in lattice QCD calculations.

The HAL QCD method is a promising method to calculate the hadron-hadron interactions in lattice QCD,
in which we construct the hadron-hadron ``potential'' from the Nambu-Bethe-Salpeter (NBS) wave function
and the physical observables such as scattering phase shifts are calculated by solving the Schr\"odinger equation
with the potential in the infinite volume~\cite{HAL:PRL2007,HAL:PTP2010,Ishii:PLB2012,HAL:PTEP2012}.
One of the unique features in the procedure of the HAL QCD method is that the spatial correlation
of the NBS wave function is calculated at all discrete points on a cubic lattice,
and thus the hadron-hadron interactions from the HAL QCD method are expected to be a good application
of Misner's method (or Misner's decomposition/extraction).
In fact, 
noticing that not all discrete points at a given radial coordinate $r$ are
necessarily transformed to each other by the cubic rotation,
it is realized that there is a room to develop a better methodology for the partial wave decomposition
than the standard projection method based on the irreducible representation of the cubic group.

In this paper, we apply Misner's method for the first time to the lattice QCD study for hadron-hadron scatterings in the framework of the HAL QCD method.
We extract the potential from
the $l=0$ ($S$-wave) component of the NBS wave function by Misner's method,
which is then compared with the conventional HAL QCD potential.
  
In the conventional method, the $S$-wave projection of the NBS wave function is approximated on the lattice by the $A_1^+$ projection as
\begin{eqnarray}
	\psi^{A_1^+} (\vec{x}) \equiv P^{A_1^+} \psi (\vec{x}) = \frac{1}{48} \sum_{g\in O_h} \psi (g^{-1}\vec{x}), \label{eq:L0projection_rot}
\end{eqnarray}
where the cubic group $O_h$ consists of cubic rotations and the parity.
The $A_1^+$ representation contains not only $l=0$ component but also higher partial waves with $l\geq4$.
If $l\geq4$ components of the NBS wave function were absent, the NBS wave function (and also the potential) would be isotropic and thus depend only on the radial coordinate $r=\vert\vec x\vert$.
However, we often observe comb-like structures in the potential in terms of $r$ (for example, see Fig.~2 in Ref.~\cite{Inoue:PTP2010}), which represent the anisotropy of the potential. 
This observation indicates the existence of non-negligible $l\geq4$ components.
We often observe that these comb-like structures lead to superficial fluctuations, whose magnitudes are larger than those of the genuine statistical fluctuations.

Throughout this paper, we write the NBS wave function as $\psi(\vec{x})$ with $\vec{x} = (x,y,z)$ or  $(r,\theta,\phi)$, which is expanded in term of the spherical harmonics $Y_{lm}(\theta,\phi)$ as
\begin{eqnarray}
	\psi(\vec{x}) = \sum_{l=0}^\infty \sum_{m=-l}^l g_{lm}(r) Y_{lm}(\theta,\phi), \label{eq:expand_lm_NBS}
\end{eqnarray}
where we call $g_{lm}(r)$ ``spherical harmonics amplitude'' for a $(l,m)$ component.

This paper is organized as follows.
After briefly reviewing the HAL QCD method in Sec.~\ref{sec:method_HAL}, we explain Misner's method in detail in Sec.~\ref{sec:method}, together with some remarks on its application to the HAL QCD method. 
Our main results are given in Sec.~\ref{sec:results}, where we consider the spin-singlet $\Lambda_c N$ system as a representative example.
In Sec.~\ref{subsec:results_NBSwave}, we extract the $l=0$ component of the NBS wave function by Misner's method, where the comb-like structures indeed disappear. 
We however found that contaminations from $l\ge 4$ partial waves are small in the  $A_1^+$ projected NBS wave function.
In Sec.~\ref{subsec:results_laplacianl}, we analyze the Laplacian of the NBS wave function by Misner's method. 
We found that $l\ge 4$ components are enhanced by applying Laplacian  to the NBS wave function
in the conventional HAL QCD method, while such a problem is absent in Misner's method, where the Laplacian is calculated analytically for each partial wave component.
In Sec.~\ref{subsec:parameter_dependence}, we investigate parameter dependencies of the potential
in Misner's method. 
In Sec.~\ref{sec:observables}, we calculate the scattering phases shifts from the HAL QCD potentials
with the conventional $A_1^+$ projection and with Misner's $S$-wave extraction.
We found that not only the central values but also statistical errors agree in both cases.
We briefly discuss a reason for this agreement. 
Summary and conclusion are presented in Sec.~\ref{sec:summary}.
In appendix~\ref{app:r-fix_Ldeco}, a simpler but less general method is considered to extract the $l=0$ component from the $A_1^+$ projected NBS wave function.
\section{HAL QCD method} 
\label{sec:method_HAL}
In the HAL QCD method~\cite{HAL:PRL2007,HAL:PTP2010,Ishii:PLB2012,HAL:PTEP2012}, a non-local but energy-independent potential is defined through the Schr\"odinger equation as 
\begin{eqnarray}
	\left( E_k - H_0 \right) \psi^{(W_k)} (\vec{x}) = \int d^3 x^\prime U (\vec{x}, \vec{x^\prime}) \psi^{(W_k)} (\vec{x^\prime}),~~
	\left(E_k = \frac{|\vec{k}|^2}{2\mu},~H_0 = -\frac{\vec{\nabla}^2}{2\mu}\right), \label{eq:Sch_eq_NBS}
\end{eqnarray}
where the NBS wave function in the center-of-mass frame is given by
\begin{eqnarray}
	\psi^{(W_k)} (\vec{x}) e^{-W_kt} = \frac{1}{\sqrt{Z_1} \sqrt{Z_2}} \sum_{\vec{y}} \langle 0 | B_1 (\vec{x} + \vec{y}, t) B_2 (\vec{y}, t) |2B; W_k \rangle .
	\label{eq:infinite_NBSwave}
\end{eqnarray}
Here we consider a two-baryon system as a representative case where $B_i (\vec{x}, t)~(i=1,2)$ is the local interpolating operator for a baryon $B_i$ with its renormalization factor $\sqrt{Z_i}$. 
The state $|2B;W_k\rangle$ stands for a QCD eigenstate for the two-baryon system at the total energy $W_k = \sqrt{|\vec{k}|^2+m_{B_1}^2}+\sqrt{|\vec{k}|^2+m_{B_2}^2}$ with baryon masses $m_{B_{1,2}}$ and a relative momentum $\vec{k}$, and $\mu$ denotes a reduced mass.
Since the asymptotic behavior of the NBS wave function at large $r = |\vec{x}|$ is identical to that of the scattering wave in quantum mechanics, whose phase shift is the phase of QCD $S$-matrix~\cite{Aoki:PRD2013,Gongyo:PTEP2018}, the potential defined from the NBS wave functions reproduces the scattering phase shifts in QCD. 
Note that the non-local potential is constructed so as to be energy-independent below the inelastic threshold~\cite{HAL:PTP2010,HAL:PTEP2012}.

In terms of the NBS wave functions, the two-baryon four-point correlation function is expressed as
\begin{eqnarray}
	G (\vec{x}, t-t_0) &=& \sum_{\vec{y}} \langle 0| B_1 (\vec{x}+\vec{y}, t) B_2 (\vec{y}, t) \mathcal{J}^{(J^P)} (t_0)| 0 \rangle \nonumber \\
	&=& \sum_n \sum_{\vec{y}} \langle 0| B_1 (\vec{x}+\vec{y}, t) B_2 (\vec{y}, t) | 2B; W_n \rangle \langle 2B; W_n | \mathcal{J}^{(J^P)} (t_0)| 0 \rangle 
	+ \cdots \nonumber \\
	&=&\sqrt{Z_1} \sqrt{Z_2} \sum_n \psi^{(W_n)} (\vec{x}) e^{-W_n (t-t_0)} A_n + \cdots, ~~
	A_n \equiv \langle 2B; W_n | \mathcal{J}^{(J^P)}(0) | 0 \rangle, ~~~~
	\label{eq:4pt_corr}
\end{eqnarray}
where $\mathcal{J}^{(J^P)} (t_0)$ stands for a source operator defined so as to create two-baryon states at $t=t_0$ with the total angular momentum $J$ and the parity $P$, and the ellipses represent contributions from inelastic states.
For a sufficiently large $t-t_0$, the NBS wave function for the ground state is extracted from the four-point function as
\begin{eqnarray}
	G (\vec{x}, t-t_0) \to \sqrt{Z_1} \sqrt{Z_2} \psi^{(W_0)} (\vec{x}) e^{-W_0 (t-t_0)} A_0 + \mathcal{O}(e^{-W_1 (t-t_0)}). \label{eq:ground_state_NBS_domination}
\end{eqnarray}

In practice, however, this extraction of the NBS wave function from the ground state saturation is very difficult due to increasing statistical noises at large $t-t_0$~\cite{Iritani:JHEP2016,Iritani:PRD2017,Iritani:JHEP2019}. 
Therefore, in Ref.~\cite{Ishii:PLB2012}, an improved method was proposed to extract the potential without the requirement of the ground state saturation.
We consider the normalized baryon four-point correlation function ($R$-correlator) defined by 
\begin{eqnarray}
	R (\vec{x}, t-t_0) &\equiv& \frac{G (\vec{x}, t-t_0)}{e^{-m_{B_1}(t-t_0)} e^{-m_{B_2}(t-t_0)}}
	= \sqrt{Z_1Z_2}\sum_n \psi^{(W_n)}(\vec{x}) e^{-\Delta W_n (t-t_0)} A_n + \cdots, \label{eq:Rcorr}
\end{eqnarray}
where $\Delta W_n = W_n - (m_{B_1} + m_{B_2})$.
If contributions from inelastic states are negligible (``elastic state saturation''), this $R$-correlator satisfies 
\begin{eqnarray}
	\left[ \left( \frac{1+3\delta^2}{8\mu} \right) \frac{\partial^2}{\partial t^2} - \frac{\partial}{\partial t} - H_0 
	+ \mathcal{O}(\frac{\delta^2}{2m_{B_1}m_{B_2}} \frac{\partial^3}{\partial t^3}) \right] R (\vec{x}, t-t_0) 
	= \int d^3 x^\prime U (\vec{x}, \vec{x^\prime}) R (\vec{x^\prime}, t-t_0), \label{eq:Sch_eq_Rcorr}
\end{eqnarray}
where $\delta = (m_{B_1}-m_{B_2})/(m_{B_1}+m_{B_2})$.
Since the elastic state saturation can be generally achieved at much smaller $t-t_0$ than the case of the ground state saturation, we can obtain reliable results from this ``time-dependent HAL QCD method''~\cite{Iritani:JHEP2016,Iritani:PRD2017,Iritani:JHEP2019,Iritani:2018zbt}. 

In order to handle the non-locality of the potential, we introduce the derivative expansion as
\begin{eqnarray}
	U(\vec{x},\vec{x^\prime}) = V(\vec{x},\vec{\nabla}) \delta^{(3)} (\vec{x}-\vec{x^\prime}), \label{eq:deriv_expansion}
\end{eqnarray}
where $V(\vec{x},\vec{\nabla})$ is expanded in terms of $\vec{\nabla}$~\cite{Okubo:AP1958}.
At low energies, since the leading order (LO) potential of the derivative expansion dominates~\cite{Iritani:2018zbt}, the interaction for the $S$-wave spin singlet system is well approximated by the LO central potential given as
\begin{eqnarray}
 	V_{\mathrm{LO}}^{{}^1S_0}(r) = \frac{\left[ \left( \frac{1+3\delta^2}{8\mu} \right) \frac{\partial^2}{\partial t^2} - \frac{\partial}{\partial t} - H_0 \right] R^{{}^1S_0} (\vec{x}, t-t_0)}{R^{{}^1S_0} (\vec{x}, t-t_0)} , \label{eq:1S0_pot_Rcorr}
\end{eqnarray}
where the $R$-correlator for the ${}^1S_0$ state is defined as
\begin{eqnarray}
	R^{{}^1S_0} (\vec{x}, t-t_0) \equiv P^{A_1^+} P^{S=0} R (\vec{x}, t-t_0). \label{eq:1S0_Rcorr}
\end{eqnarray}
Here $P^{S=0}$ represents the projection to the state with the total spin $S=0$, while $P^{A_1^+}$ stands for the projection to the $A_1^+$ representation in Eq.~\eqref{eq:L0projection_rot}.
As we have explained before, the $A_1^+$ projection contains not only $l=0$ component but also $l\ge4$ components, which produces angular dependencies of the $R$-correlator as well as the potential.
\section{Approximated partial wave decomposition} \label{sec:method}
\subsection{Misner's method} \label{subsec:method_misner}
Let us consider the extraction $g_{lm}(r)$ for a given $r\equiv\vert \vec x\vert =R$ from the NBS wave function.
In the continuum space, we can obtain $g_{lm}(R)$ by taking the spherical surface integral at $r=R$ as
\begin{eqnarray}
	g_{lm}(R) = \int_S d\Omega~\overline{Y_{lm}(\theta,\phi)} \psi(\vec{x}; r=R), \label{eq:sph_int_g00_continuum}
\end{eqnarray}
because of the orthogonality of the spherical harmonics,
\begin{eqnarray}
	\int_S d\Omega~\overline{Y_{lm}(\theta,\phi)} Y_{l^\prime m^\prime}(\theta,\phi) = \delta_{ll^\prime} \delta_{mm^\prime},  \label{eq:sph_orthogonality}
\end{eqnarray}
where the overline represents its complex conjugation.
This is even true on a finite $L_s^3$ box in the continuum space for $0 < R \le L_s/2$.

In the discrete space such as the cubic lattice, however, we can not obtain $g_{lm}(R)$ for any $R$, since $N_R$, the number of points which satisfy $r =R$, is finite, so that the infinite dimensional matrix,
\begin{eqnarray}
	{\cal G} \equiv \left\{ \left. {\cal G}_{lm,l^\prime m^\prime} \equiv \frac{1}{N_R} \sum_{\vec{x}\in\{\vec{x} | r=R\}}
	\overline{Y_{lm}(\theta,\phi)} Y_{l^\prime m^\prime}(\theta,\phi) \right\vert \vert m\vert \le l,   \vert m^\prime\vert \le  l^\prime \right\},
\end{eqnarray}
is non-invertible due to its zero eigenvalues.
If one knows that contributions from higher partial waves are negligible, one can introduce some approximation in order to obtain $g_{lm}(R)$ for small $l$ and some restricted $R$.
For example, we may impose the condition that $l,l^\prime \le l_{\rm max}$, where $l_{\rm max}$ is chosen so that the finite dimensional matrix ${\cal G}^{l_{\rm max}} \equiv \{{\cal G}_{lm,l^\prime m^\prime} \vert l,l^\prime \le l_{\rm max} \}$ becomes invertible. 
In appendix~\ref{app:r-fix_Ldeco}, we consider this approximation in detail.

In Ref.~\cite{Misner:CQG2004}, a more general and sophisticated approximation was proposed.
To explain the method, let us start from the continuum case.
We first introduce the basis functions in the radial coordinate $G^{R,\Delta}_n(r)$ ($n=0,\cdots,\infty$) which are orthonormal in  the radial interval $[R-\Delta,R+\Delta]$ as
\begin{eqnarray}
	\int_{R-\Delta}^{R+\Delta} dr~r^2~\overline{G^{R,\Delta}_n(r)} G^{R,\Delta}_m(r) = \delta_{nm} .\label{eq:orthonormal_basis_r}
\end{eqnarray}
One of the candidates for $G^{R,\Delta}_n(r)$ is given by
\begin{eqnarray}
	G^{R,\Delta}_n(r) = \frac{1}{r} \sqrt{\frac{2n+1}{2\Delta}} P_n \left( \frac{r-R}{\Delta} \right), \label{eq:orthonormal_basis_r_Legendre}
\end{eqnarray}
where $P_n(r)$ is the Legendre polynomial, which obviously satisfies Eq.~\eqref{eq:orthonormal_basis_r}.
When we consider a spherical shell $S_{R,\Delta}$ with thickness $2\Delta$ surrounding the sphere surface $r=R$ defined by
\begin{eqnarray}
	S_{R,\Delta} \equiv \left\{ \vec{x} | R-\Delta \leq r \leq R+\Delta \right\}, \label{eq:sph_shell_def}
\end{eqnarray}
an orthonormal basis function $\mathcal{Y}^{R,\Delta}_{nlm}(r,\theta,\phi) \equiv G^{R,\Delta}_n(r) Y_{lm}(\theta,\phi)$ obeys
\begin{eqnarray}
	\int_{S_{R,\Delta}} d^3x~\overline{\mathcal{Y}^{R,\Delta}_{nlm}(r,\theta,\phi)} \mathcal{Y}^{R,\Delta}_{n^\prime l^\prime m^\prime}(r,\theta,\phi) = \delta_{nn^\prime}\delta_{ll^\prime}\delta_{mm^\prime}, \label{eq:orthonormal_basis_rsp}
\end{eqnarray}
where the integral over $S_{R,\Delta}$ is defined as
\begin{eqnarray}
	\int_{S_{R,\Delta}} d^3 x \equiv \int_{R-\Delta}^{R+\Delta} r^2 dr \int_S d\Omega. \label{eq:integral_S_def}
\end{eqnarray}
The NBS wave function in a spherical shell $S_{R,\Delta}$ is expanded in terms of the orthonormal basis functions as 
\begin{eqnarray}
	\psi(\vec{x}) = \sum_{n=0}^\infty \sum_{l=0}^\infty \sum_{m=-l}^l c^{R,\Delta}_{nlm}~\mathcal{Y}^{R,\Delta}_{nlm}(r,\theta,\phi) \label{eq:expand_nlm_NBS}
\end{eqnarray}
with coefficients $c^{R,\Delta}_{nlm}$, which can be determined by
\begin{eqnarray}
	c^{R,\Delta}_{nlm} = \int_{S_{R,\Delta}} d^3x~\overline{\mathcal{Y}^{R,\Delta}_{nlm}(r,\theta,\phi)}~\psi(\vec{x}). \label{eq:sph_int_gLM_continuum}
\end{eqnarray}
We finally obtain the spherical harmonics amplitude $g_{lm}(r)$ for $ R -\Delta \le r \le R+\Delta$ as
\begin{eqnarray}
	g_{lm}(r) = \sum_{n=0}^\infty~c^{R,\Delta}_{nlm}~G^{R,\Delta}_n(r) .\label{eq:gLM_from_Cnlm_continuum}
\end{eqnarray}

Ref.~\cite{Misner:CQG2004} employed ${\cal Y}^{R,\Delta}_{n,l,m}(r,\theta,\phi)$ as the basis function for the approximation in the case of the discrete space on a cubic lattice.
In this case, the volume integral is replaced by the discrete sum as
\begin{eqnarray}
	\int_{S_{R,\Delta}} d^3x~\Longrightarrow~\sum_{\vec{x}} \omega^{R,\Delta}({\vec{x}}), \label{eq:sph_int_lattice}
\end{eqnarray}
where $\omega^{R,\Delta}({\vec{x}})$ is a weight factor, which corresponds to a volume of the overlapped region between the shell $S_{R,\Delta}$ and a unit cube around the point $\vec{x}$.
For example, if the unit cube lies entirely inside the shell $S_{R,\Delta}$, $\omega^{R,\Delta}({\vec{x}}) = a^3$ with a lattice spacing $a$, while $\omega^{R,\Delta}({\vec{x}}) = 0$ when the unit cube lies entirely outside the shell.
Since $\omega^{R,\Delta}({\vec{x}})$ in the general cases is rather complicated, it is approximated in Ref.~\cite{Misner:CQG2004} as 
\begin{eqnarray}
	\omega^{R,\Delta}({\vec{x}}) = 
        \begin{cases}
		a^3 & \text{for $|r-R| < \Delta - \frac{1}{2} a$}, \\
		0   & \text{for $|r-R| > \Delta + \frac{1}{2} a$},\\
		a^2 \left( \Delta + \frac{1}{2} a - |R-r| \right)
                & \text{otherwise,}
	\end{cases}
	\label{eq:misner_weight_def}
\end{eqnarray}
which corresponds to the overlapped volume of a unit cube parallel to the radial direction.
Using this, we define an inner product of functions $f(\vec{x})$ and $g(\vec{x})$ in the shell $S_{R,\Delta}$ as
\begin{eqnarray}
	\langle f | g \rangle_{S_{R,\Delta}} &\equiv& \sum_{\vec{x}} \omega^{R,\Delta}({\vec{x}})~\overline{f(\vec{x})}~g(\vec{x}). \label{eq:inner_prod_lattice}
\end{eqnarray}
Let us consider ${\cal G}_{AA^\prime} \equiv \langle {\cal Y}_A^{R,\Delta} \vert   {\cal Y}_{A^\prime}^{R,\Delta} \rangle_{S_{R,\Delta}}$ with a shorthand notation $A=(n,l,m)$.
The finite dimensional Hermitian matrix ${\cal G}$ constructed from ${\cal G}_{AA^\prime}$ with a restriction that $n,n^\prime \le n_{\rm max}$ and $l,l^\prime \le l_{\rm max}$ becomes invertible if one properly chooses $n_{\rm max}$ and $l_{\rm max}$.
Using ${\cal G}$ (whose dependencies on $n_{\rm max}$ and $l_{\rm max}$ are implicit here), one can define the dual basis functions $\tilde{\mathcal{Y}}^{R,\Delta}_A$ as
\begin{eqnarray}
	\tilde{\mathcal{Y}}^{R,\Delta}_A(\vec{x}) \equiv {\sum_{B}}^\prime \mathcal{Y}^{R,\Delta}_B(\vec{x})~\mathcal{G}^{-1}_{BA}, 
	\label{eq:adjoint_basis_def}
\end{eqnarray}
which satisfies
\begin{eqnarray}
	\langle \tilde{\mathcal{Y}}^{R,\Delta}_A | \mathcal{Y}^{R,\Delta}_B \rangle_{S_{R,\Delta}} &=& {\sum_{C}}^\prime \mathcal{G}_{AC}^{-1} \langle \mathcal{Y}^{R,\Delta}_C | \mathcal{Y}^{R,\Delta}_B \rangle_{S_{R,\Delta}} = {\sum_C}^\prime \mathcal{G}_{AC}^{-1} \mathcal{G}_{CB} = \delta_{AB}, \label{eq:ampl_const_lattice}
\end{eqnarray}
where the prime in the summation indicates the upper bounds $n_{\rm max}$ and $l_{\rm max}$.

Assuming that $c^{R,\Delta}_{nlm}$ is negligibly small for  $l > l_{\rm max}$ or $n > n_{\rm max}$, Eq.~\eqref{eq:expand_nlm_NBS} is approximately written as 
 \begin{eqnarray}
	\psi(\vec{x}) \simeq \sum_{n=0}^{n_{\rm max}} \sum_{l=0}^{l_{\rm max}} \sum_{m=-l}^l c^{R,\Delta}_{nlm}~\mathcal{Y}^{R,\Delta}_{nlm}(r,\theta,\phi)
	\label{eq:expand_nlm_NBS_lat} 
\end{eqnarray}
with the coefficient $c^{R,\Delta}_{nlm} = \langle \tilde{\mathcal{Y}}^{R,\Delta}_{nlm} \vert \psi \rangle_{S_{R,\Delta}}$.
Finally the spherical harmonics amplitude $g_{lm}(R)$ can be approximated as
\begin{eqnarray}
	g_{lm} (R) &\simeq & \sum_{n=0}^{n_{\rm max}} c_{nlm}^{R,\Delta} G_n^{R,\Delta}(R) .
\end{eqnarray}
\subsection{Misner's method as a minimization} \label{sec:least_square}
Misner's method can be also understood from the viewpoint of the least square minimization and we here give the explicit correspondence following Ref.~\cite{Rupright:2006}.

Let us denote a $N_{R,\Delta}$ component vector $\bm{\Psi}$ of the NBS wave function as
\begin{eqnarray}
	\bm{\Psi} &=&
	\left(
	\begin{array}{c}
	\psi(\vec x_1) \\
	\psi(\vec x_2) \\
	\vdots \\
	\psi(\vec x_{N_{R,\Delta}}) \\
	\end{array}
	\right),
\end{eqnarray}
where $N_{R,\Delta}$ is the number of points in the shell $S_{R,\Delta}$, equivalently, the number of data with non-zero $\omega^{R,\Delta}(\vec{x})$.
Similarly, we define a $N_{R,\Delta}\times M$ rectangular matrix  {\boldmath$Y$} of the basis functions, whose components are defined by
\begin{eqnarray}
	\mbox{\boldmath$Y$}_{i,nlm} = \mathcal{Y}^{R,\Delta}_{n,l,m}(\vec{x}_i), \quad 
	1\le i \le N_{R,\Delta},  \ 0\le n \le n_{\rm max}, \
	0\le l \le l_{\rm max},\ \vert m\vert \le l, 
\end{eqnarray}
where $M$ is the number of the basis functions and is given by $M = (n_{\rm max} +1)  (l_{\rm max} +1) ^2$.
For $n_{\mathrm{max}} = 2$ and $l_{\mathrm{max}} = 2$, for example, {\boldmath$Y$} becomes
\begin{eqnarray}
	\mbox{\boldmath$Y$} =
	\left(
	\begin{array}{ccccc}
		\mathcal{Y}^{R,\Delta}_{0,0,0}(\vec{x}_1) & \mathcal{Y}^{R,\Delta}_{0,1,-1}(\vec{x}_1) & \mathcal{Y}^{R,\Delta}_{0,1,0} (\vec{x}_1) & \cdots & \mathcal{Y}^{R,\Delta}_{2,2,2} (\vec{x}_1) \\
		\vdots & \vdots & \vdots &  & \vdots \\
		\mathcal{Y}^{R,\Delta}_{0,0,0}(\vec{x}_{N_{R,\Delta}}) & \mathcal{Y}^{R,\Delta}_{0,1,-1}(\vec{x}_{N_{R,\Delta}}) & \mathcal{Y}^{R,\Delta}_{0,1,0} (\vec{x}_{N_{R,\Delta}}) & \cdots & \mathcal{Y}^{R,\Delta}_{2,2,2} (\vec{x}_{N_{R,\Delta}})
	\end{array}
	\right),
\end{eqnarray}
where the number of columns is $M =  27$.
Defining a $N_{R,\Delta}\times N_{R,\Delta}$ diagonal matrix {\boldmath$W$} for the non-zero weight $\omega^{R,\Delta}(\vec{x})$, the $M\times M$ matrix $\bm{\mathcal{G}}\equiv\{ \mathcal{G}_{AB} \}$ is simply written as $\bm{\mathcal{G}} = $ {\boldmath$Y$}${}^\dagger${\boldmath$ W Y$}.

Using these notations, we introduce a trial $N_{R,\Delta}$-component vector function $\bm{\tilde\Psi} \equiv ${\boldmath $Y \tilde C$} as
\begin{eqnarray}
	\bm{\tilde\Psi}_i &\equiv & \tilde \psi(\vec x_i) = \sum_{n=0}^{n_{\rm max}} 
	\sum_{l=0}^{l_{\rm max}} \sum_{m=-l}^l  \mbox{\boldmath$Y$}_{i,nlm} \mbox{\boldmath$\tilde C$}_{nlm},  
\end{eqnarray}
where a $M$-component vector {\boldmath$\tilde C$} corresponds to fit parameters that should minimize
\begin{eqnarray}
	F(\mbox{\boldmath$\tilde C$} ) &=& \left( \tilde{\bm{\Psi}}-\bm{\Psi} \right)^\dagger \mbox{\boldmath$ W$}
	\left( \tilde{\bm{\Psi}}-\bm{\Psi} \right) \nonumber 
	= \left( \mbox{\boldmath$Y \tilde C$}-\bm{\Psi} \right)^\dagger \mbox{\boldmath$ W$} \left( \mbox{\boldmath$Y \tilde C$}-\bm{\Psi} \right). 
	\label{eq:weighted_sum_square_residue}
\end{eqnarray}
Since $dF(\mbox{\boldmath$\tilde C$})/d\mbox{\boldmath$\tilde C$} = 0$ at the minimum $\mbox{\boldmath$\tilde C$} = \mbox{\boldmath$\tilde C$}_{\rm min}$, we obtain
\begin{eqnarray}
	\mbox{\boldmath$\tilde C$}_{\mathrm{min}} &=& \left( \mbox{\boldmath$Y$}^\dagger \mbox{\boldmath$W Y$} \right)^{-1} \mbox{\boldmath$Y$}^\dagger \mbox{\boldmath$W$}~\bm{\Psi} \nonumber 
	= \bm{\mathcal{G}}^{-1} \mbox{\boldmath$Y$}^\dagger \mbox{\boldmath$W$}~\bm{\Psi},\label{eq:min_weighted_residue_solution}
\end{eqnarray}
so that 
\begin{eqnarray}
	\left(\bm{\tilde\Psi}_{\rm min}\right)_i
	&=&\left( \mbox{\boldmath$Y $} \mbox{\boldmath$\tilde C$}_{\mathrm{min}}\right)_i =
	{\sum_A}^\prime
	{\cal Y}_{A}^{R,\Delta}(\vec x_i)  \langle \tilde{\cal Y}_{A}^{R,\Delta} \vert \psi\rangle_{S_{R,\Delta}} 
	= \sum_{n=0}^{n_{\rm max}} \sum_{l=0}^{l_{\rm max}}\sum_{m=-l}^l  
	{\cal Y}_{nlm}^{R,\Delta}(\vec x_i) c_{nlm}^{R,\Delta} ,~~~~
\end{eqnarray}
which agrees with Misner's method, Eq.~(\ref{eq:expand_nlm_NBS_lat}).
Therefore, Misner's method is equivalent to finding a solution of $\tilde\psi(\vec x_i)$ which minimizes the difference between the data $\psi(\vec x_i)$ and the fit function $\tilde\psi(\vec x_i)$ defined by the norm $\langle \tilde \psi -\psi \vert  \tilde \psi -\psi \rangle_{S_{R,\Delta}}$.
\subsection{Remarks}
\label{subsec:remarks}
The calculation of the potential in the HAL QCD method requires the Laplacian applied to the NBS wave function, which is conventionally approximated by a finite difference, and thus contains discretization errors.
In the application of Misner's decomposition to the HAL QCD method, we can instead employ an analytical expression for the Laplacian, which operates on the (approximately obtained) partial wave $g_{lm}(r) Y_{lm}(\theta,\phi)$ as
\begin{eqnarray}
	\vec{\nabla}^2\left[ g_{lm}(r) Y_{lm}(\theta,\phi)\right] &=& \left\{\frac{1}{r} \frac{\partial^2}{\partial r^2} \left[ rg_{lm}(r) \right] -\frac{l(l+1)}{r^2} g_{lm}(r)\right\} Y_{lm}(\theta, \phi), \label{eq:Laplacian_sph_harm_Mis0}
\end{eqnarray}
where
\begin{eqnarray}
	\frac{1}{r} \frac{\partial^2}{\partial r^2} \left[ rg_{lm}(r) \right] &=& \sum_{n=0}^{n_{\rm max}}~c^{R,\Delta}_{nlm}~\frac{1}{r} \frac{\partial^2}{\partial r^2} \left[ rG^{R,\Delta}_n(r) \right] \nonumber \\ &=&
	\frac{1}{r} \sum_{n=0}^{n_{\rm max}}~c^{R,\Delta}_{nlm}~\frac{\partial^2}{\partial r^2} \left[ \sqrt{\frac{2n+1}{2\Delta}} P_n \left( \frac{r-R}{\Delta} \right)\right] \nonumber \\&=& 
	\frac{1}{r\Delta^2}~\sum_{n=0}^{n_{\rm max}}~\sqrt{\frac{2n+1}{2\Delta}}~c^{R,\Delta}_{nlm}~P^{\prime\prime}_n \left( \frac{r-R}{\Delta} \right), 
	\label{eq:Laplacian_sph_harm_Mis}
\end{eqnarray}
and $P^{\prime\prime}_n$ is the second-order derivative for the Legendre polynomial.
Unlike the conventional HAL QCD method
in which the difference operator for the Laplacian
is applied to (all partial wave components of) the NBS wave function,
it is clear that the analytic derivative in Misner's method does not induce contributions from other partial waves than the targeted one. 
A comparison between two implementations for the Laplacian operator will be given in Sec.~\ref{subsec:results_laplacianl}.

In Misner's method, it is practically important to choose $n_{\rm max}$, $l_{\rm max}$ and $\Delta$ appropriately.
While larger $n_{\rm max}, l_{\rm max}$ and smaller $\Delta$ gives a better approximation of the NBS wave function, it leads to a small $N_{R,\Delta} - M$ that may cause some numerical instability due to small eigenvalues of $\mathcal{G}_{AB}$\footnote{If $N_{R,\Delta}-M <0$ (and there is no symmetry), $\mathcal{G}_{AB}$ has zero eigenvalues.} or may give an over-fitting, where $N_{R,\Delta}$ and $M$ correspond to the numbers of data and fit parameters, respectively. 
For example, if the variation of the NBS wave function in the radial coordinate is large, one should increase $n_{\mathrm{max}}$ to approximate the spherical harmonics amplitude $g_{lm}(r)$ better, but not too much so as to avoid the instability or the over-fitting.
In Ref.~\cite{Fiske:CQG2006}, the scaling of the discretization error in  Misner's method is discussed.
  It is found that the error depends on $\Delta$ as ${\cal O}(\Delta^{n_{\mathrm{max}}+2})$,
  which also indicates that the choice of $\Delta = {\cal O}(a)$ is preferable.
  In addition, the volume integral in the shell with the approximated weight (Eq.~(\ref{eq:misner_weight_def}))
  gives ${\cal O}(a^2)$ error with the choice of $\Delta = {\cal O}(a)$.
  Therefore, the choice of the parameters $n_{\mathrm{max}} = 2$ with $\Delta={\cal O}(a)$
  is found to be good for the second order accuracy.

The author~\cite{Fiske:CQG2006} also suggests $\Delta=3a/4$ as a rule of thumb by numerical investigations, but we have to examine whether the results are stable against the change of parameters $\Delta, n_{\mathrm{max}}, l_{\mathrm{max}}$ case by case, as will be presented in the next section. 

In practice, the most costly calculation in Misner's method is the construction of the matrix $\mathcal{G}_{AB}$.
Once we calculate the matrix, however, we can use it for different lattice data (e.g. NBS wave functions calculated on different gauge samples).
Therefore, it is better to calculate the dual basis functions (Eq.~\eqref{eq:adjoint_basis_def}) once before the calculation of the spherical harmonics amplitude from NBS wave functions and use them for  the later analyses.
One possible obstacle in this procedure is that the dual basis functions {$\tilde{\mathcal{Y}}^{R,\Delta}_{nlm}(\vec{x})$} consume large amount of memory to store,  $L^3 (n_{\rm max}+1) (l_{\rm max} +1)^2\times 16$ Bytes for each given value of $R$.
For example, the required memory for $\tilde{\mathcal{Y}}^{R,\Delta}_{nlm}(\vec{x})$ with $L=32$, $n_{\mathrm{max}} = 4$ and $l_{\mathrm{max}} = 6$ becomes $32^3 \times 5 \times 7^2 \times 16$ (Bytes) $= 122.5$ MB.
In order to reduce the memory usage by a factor of $(n_{\rm max}+1)$, we instead store
\begin{eqnarray}
	F^{R,\Delta}_{lm}(\vec{x}) \equiv \sum_{n=0}^{n_\mathrm{max}} \overline{G^{R,\Delta}_n(R)}~\tilde{\mathcal{Y}}^{R,\Delta}_{nlm}(\vec{x}) ,
\end{eqnarray}
which needs only $24.5$ MB.
Using this function, the spherical harmonics amplitude $g_{lm}(R)$ can be calculated directly as
\begin{eqnarray}
	\langle F^{R,\Delta}_{lm} | \psi \rangle_{S_{R,\Delta}} &=& \sum_{n=0}^{n_\mathrm{max}} G^{R,\Delta}_n(R)~\langle \tilde{\mathcal{Y}}^{R,\Delta}_{nlm} | \psi \rangle_{S_{R,\Delta}} 
	= \sum_{n=0}^{n_\mathrm{max}} G^{R,\Delta}_n(R)~c^{R,\Delta}_{nlm} 
	\simeq g_{lm}(R).
\end{eqnarray}
Furthermore, it is sufficient to store $F^{R,\Delta}_{lm}(\vec{x})$  only at point $\vec x$ included in the shell $S_{R,\Delta}$ where the weight function $\omega^{R,\Delta}({\vec{x}})$ is non-zero, which leads to extra large reduction for the memory usage.
\section{HAL QCD potentials with Misner's method} \label{sec:results}
For the numerical calculation, we consider the spin-singlet $\Lambda_cN$ system in the $(2+1)$-flavor full lattice QCD with the renormalization-group improved Iwasaki gluon action and a nonperturbatively ${\cal O}(a)$ improved Wilson-clover quark action on a $(32a~\mathrm{fm})^3\times(64a~\mathrm{fm})$ volume with the lattice spacing $a\simeq0.0907$ fm at $m_\pi\simeq700$ MeV.
We apply Misner's method to the same data of NBS wave function calculated in Ref.~\cite{Miyamoto:NPA2018}, where the results in the conventional HAL QCD method are given. 
For the source operator, we employ the wall-type source operator and thus the $\Lambda_cN$ system belongs to the $A_1^+$ representation of the cubic group.
In order to reduce the statistical fluctuations, we also impose the $A_1^+$ projection on the sink operator as given in Eq.~(\ref{eq:L0projection_rot}).
Recall that the $A_1^+$ representation contains the partial waves $l = 0, 4, 6, \cdots$.
In this study, we consider the $A_1^+$ projected $R$-correlator  taken at  $t-t_0=13a$.
The total number of configuration is 399, and the statistical errors are estimated by the jackknife method with a bin size of 57 configurations (the total number of bins is 7).
For more details on the simulation setup, see Ref.~\cite{Miyamoto:NPA2018}.
\subsection{NBS wave function for the spin-singlet $\Lambda_c N$ system} \label{subsec:results_NBSwave}
\begin{figure}[htb] \centering
	\includegraphics[width=16cm]{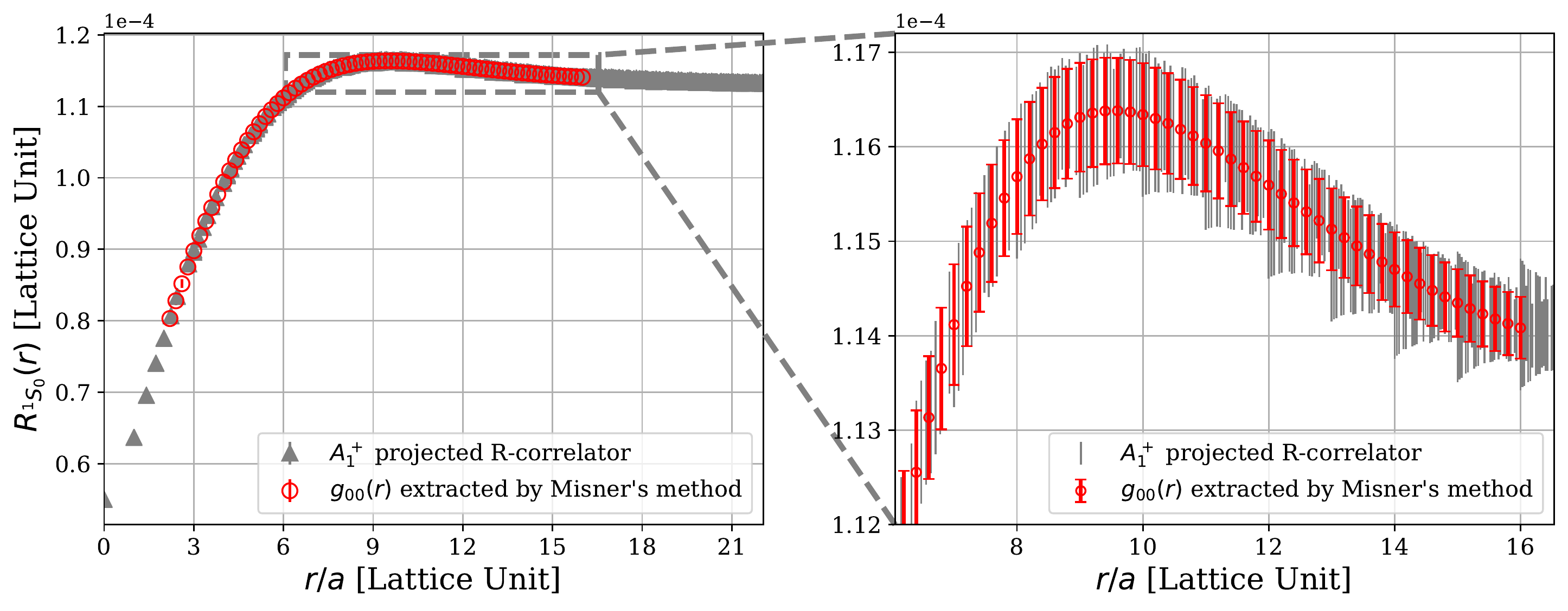}
	\caption{The $R$-correlator for spin-singlet $\Lambda_c N$ system at $t-t_0 = 13a$ for $m_{\pi}\simeq $~700 MeV. 
	The gray points show the $R$-correlator with the $A_1^+$ projection divided by $Y_{00}$, while the red points correspond to the spherical harmonics amplitude $g_{00}(r)$ calculated by Misner's method. 
\label{fig:LcN_Rcor_1S0_t13}}
\end{figure}
\begin{figure}[htb] \centering
	\includegraphics[width=16cm]{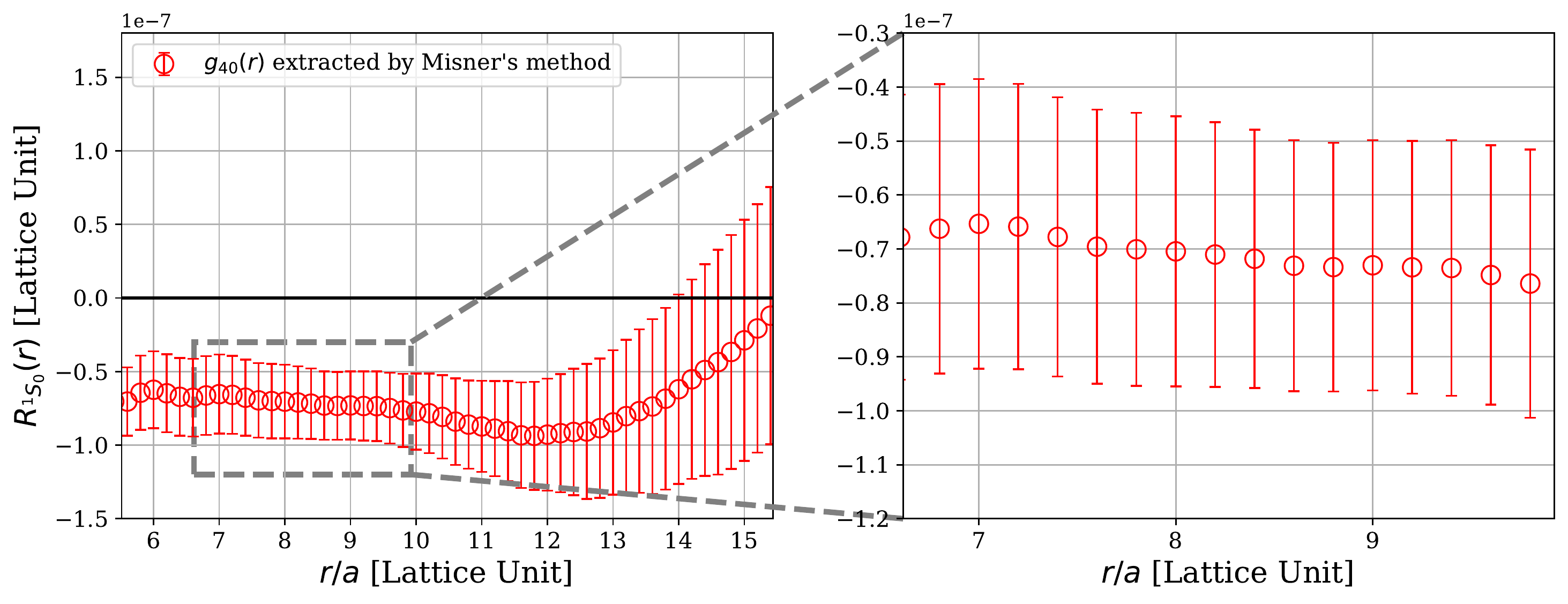}
	\caption{The spherical harmonics amplitudes for the $l=4$ component extracted by Misner's method. \label{fig:LcN_Rcor_1S0_t13_L4}}
\end{figure}
Fig.~\ref{fig:LcN_Rcor_1S0_t13} shows the results of the $R$-correlator defined in Eq.~\eqref{eq:Rcorr} for spin-singlet $\Lambda_c N$ system. 
The gray points represent the $A_1^+$ projected $R$-correlator divided by $Y_{00}$, which is actually used in Ref.~\cite{Miyamoto:NPA2018} to construct the $\Lambda_cN$ potential.
The red points correspond to the spherical harmonics amplitude $g_{lm}(r)$ for $l=m=0$ component calculated by Misner's method for the radial coordinate $r$ from $2a$ to $16a$ with the interval $\Delta r=0.2a$, so that some data are used several times. 
We do not perform Misner's method for $r < 2a$ and $r > L_s/2 = 16a$:
In the former case, the number of data points in the spherical shell $S_{r,\Delta}$ is too small, whereas the spherical shell is not contained in the $L_s^3$ cubic lattice for the latter.
We here employ $\Delta=a$, $n_{\mathrm{max}}=2$ and $l_{\mathrm{max}}=4$ as the parameters in Misner's method, and we found that $g_{00}(r)$ has a weak parameter dependence. 

Fig.~\ref{fig:LcN_Rcor_1S0_t13} shows small comb-like structures in the $A_1^+$ projected $R$-correlator, which however do not appear in the $l=0$ component extracted by Misner's method. 
This observation indicates that $l\ge4$ components exist in the $A_1^+$ projected $R$-correlator and their angular dependencies become manifest as the comb-like structures in the radial-coordinate.
Such higher partial wave components can be explicitly extracted by Misner's method as shown in Fig.~\ref{fig:LcN_Rcor_1S0_t13_L4} for the $l=4$ component ($g_{40}(r)=g_{4\:\pm4}(r)$).
Note that $g_{4m}(r)=0$ for $m \neq 0, \pm4$ for the $A_1^+$ representation.
We find that $l=4$ component indeed exists while its magnitude is small (by a factor of ${\cal O}(10^{-3})$ compared to that of the $l=0$ component).
On the other hand, the absence of comb-like structures in $l=0, 4$ components obtained by Misner's method with $l_{\mathrm{max}}=4$ indicates that $l\le 4$ components are sufficient to explain the $A_1^+$ projected $R$-correlator, which is explicitly confirmed by observing that $l=6$ component extracted by Misner's method with $l_{\mathrm{max}}=6$ is actually negligible.

  The mixing of $l=4$ component is most likely induced due to the
  rotational symmetry breaking by the finite volume (IR-effect),
  except for $r \lesssim a$ where there could also exist the effect by the finite lattice spacing (UV-effect).

\subsection{Laplacian and HAL QCD potential} 
\label{subsec:results_laplacianl}
\begin{figure}[htb] \centering
	\includegraphics[width=16cm]{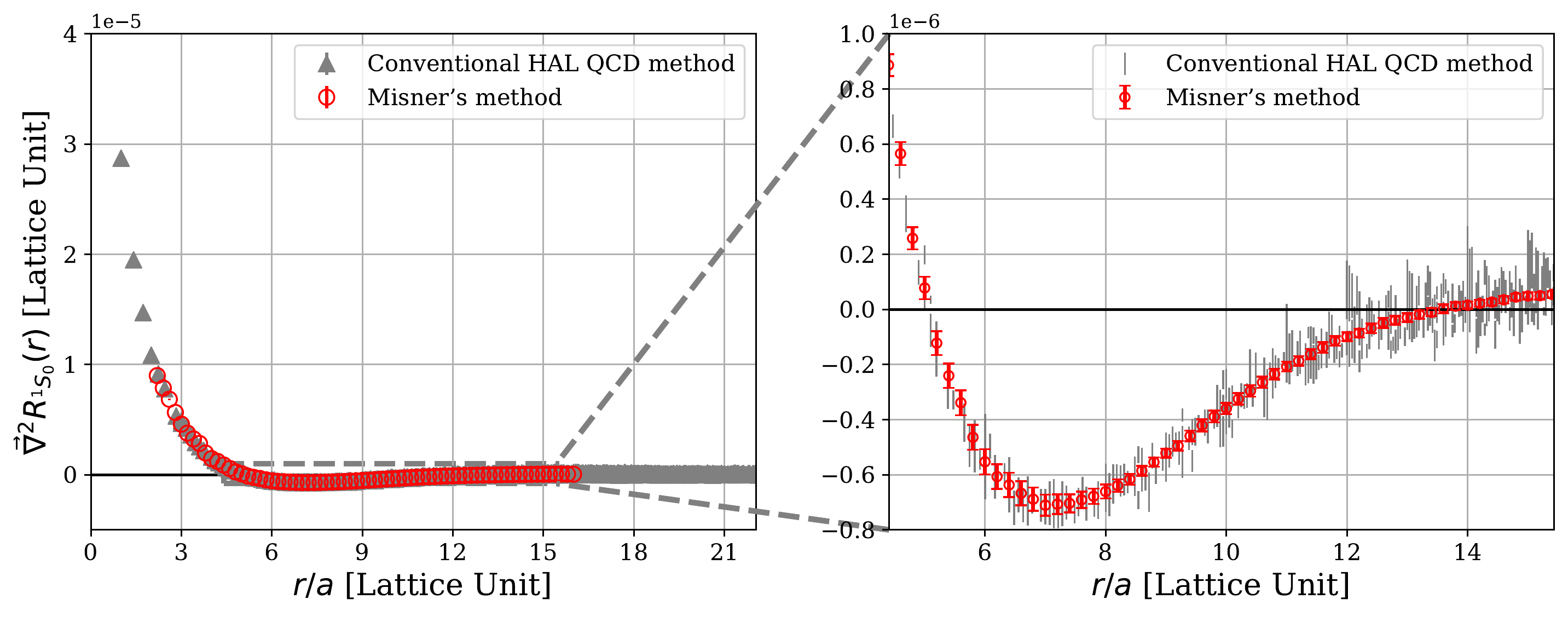} 
	\caption{The Laplacian applied to the $R$-correlator in the conventional method (gray points) and in the Misner's method (red points).
	The Laplacian is defined by a finite second-order difference in the former, while it is analytically calculated using Eq.~\eqref{eq:Laplacian_sph_harm_Mis0} in the latter. \label{fig:LcN_Lap_1S0_t13}}
\end{figure}
\begin{figure}[htb] \centering
	\includegraphics[width=16cm]{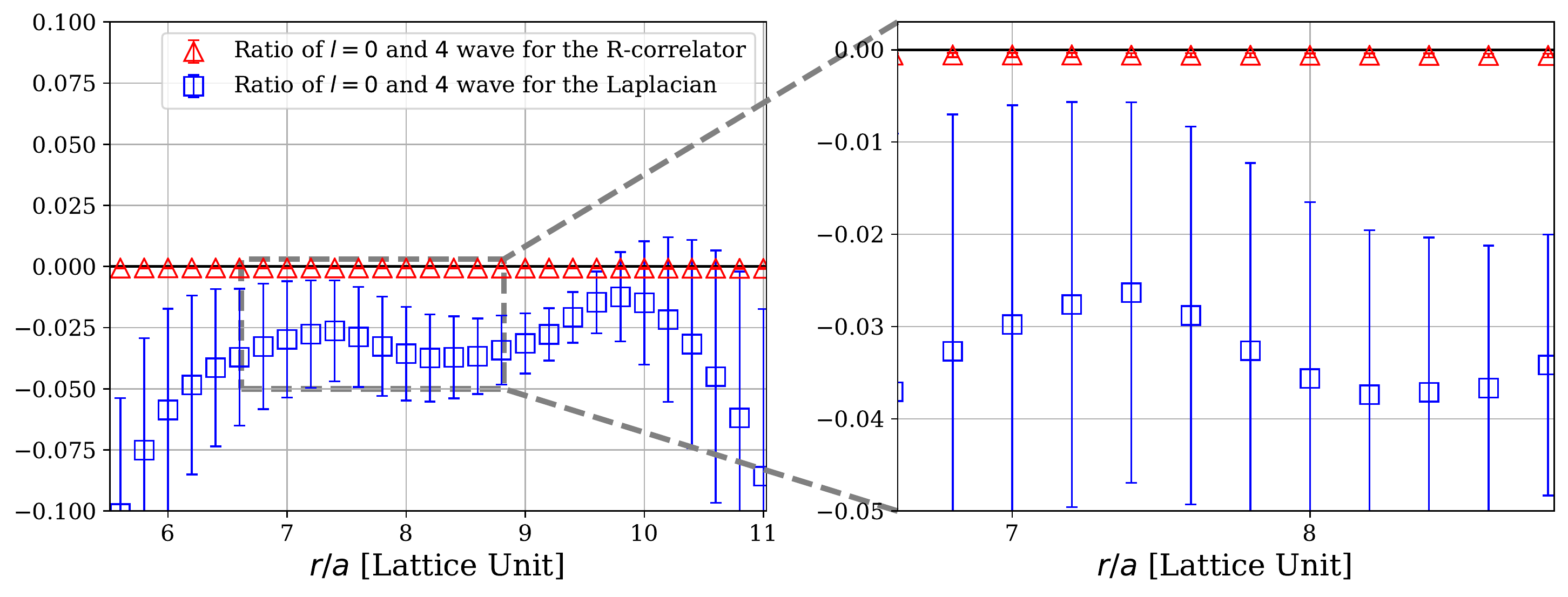}
	\caption{Ratio of the spherical harmonics amplitudes, $g_{40}(r)/g_{00}(r)$,
	for the $R$-correlator (red triangles) and  for the Laplacian term (blue squares). \label{fig:LcN_Lap_1S0_t13_L4}}
\end{figure}
We then study the effect of Laplacian applied to the NBS wave function.
Note that the term containing the Laplacian, i.e., the third term in the rhs of Eq.~(\ref{eq:1S0_pot_Rcorr}), is known to give the dominant contribution for the potential.
In the conventional HAL QCD method, the Laplacian  approximated by a finite second-order difference
is applied to (all partial wave components of) the NBS wave function, while it is analytically calculated as Eq.~\eqref{eq:Laplacian_sph_harm_Mis0} in Misner's method for the designated partial wave component.
In Fig.~\ref{fig:LcN_Lap_1S0_t13}, we compare the Laplacian applied to the NBS wave functions between Misner's method (red points) and the conventional HAL QCD method (gray points).
In the case of Misner's method, the Laplacian applied to $g_{00}(r)$ analytically does not exhibit the comb-like structure. 
In the case of the conventional HAL QCD method, on the other hand, we find that the comb-like structures are much larger than those of the $R$-correlator itself, indicating that the $l\ge4$ components are larger for the conventional Laplacian.
The partial wave decomposition of the  conventional Laplacian term reveals that the $l=4$ component is indeed
larger than the case of the $R$-correlator as shown in Fig.~\ref{fig:LcN_Lap_1S0_t13_L4}.
  The origin of these $l\geq 4$ components in the conventional Laplacian
  is most likely the $l\geq 4$ components in the $R$-correlator enhanced by the Laplacian,
  rather than the discretization error in the conventional Laplacian operator itself.
  In fact, 
  in the case of Misner's method,
  the difference between the results with the conventional Laplacian operator applied to $g_{00}(r)$
  and those with the analytic Laplacian operator
  is found to be marginal.

\begin{figure}[htb] \centering
	\includegraphics[width=16cm]{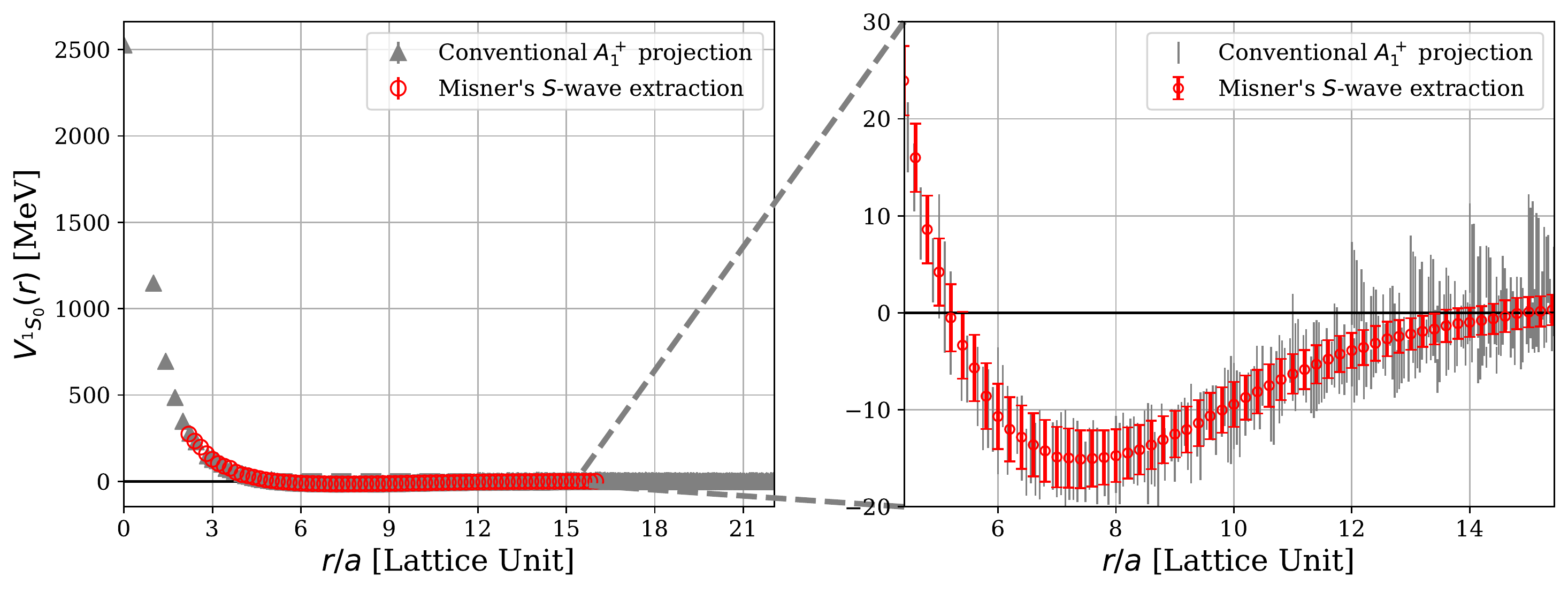} 
	\caption{The potential for the spin-singlet $\Lambda_c N$ system at $m_{\pi}\simeq $~700 MeV,  constructed by the time-dependent HAL QCD method  at $t-t_0 = 13a$. 
	  The gray points show the potential from the conventional $A_1^+$ projection, while the red points corresponds to
    that from Misner's $S$-wave extraction. \label{fig:LcN_potential_1S0_t13}
  }
\end{figure}
Shown in Fig.~\ref{fig:LcN_potential_1S0_t13} are the HAL QCD potentials for the spin-singlet $\Lambda_c N$ system 
from Misner's $S$-wave extraction (red points) and the conventional $A_1^+$ projection (gray points).
In the case of Misner's extraction, the potential is found to be free from comb-like structures.
In the case of the conventional projection, however, the potential has large comb-like structures.
The main origin is attributed to $l\ge4$ components in the Laplacian term for the potential,
which are enhanced by applying the Laplacian to the $R$-correlator, even though $l\ge4$ components are small in the $R$-correlator itself.

\subsection{Parameter dependencies for potentials in Misner's method}
\label{subsec:parameter_dependence}
\begin{figure}[htbp] \centering
	\includegraphics[width=16cm]{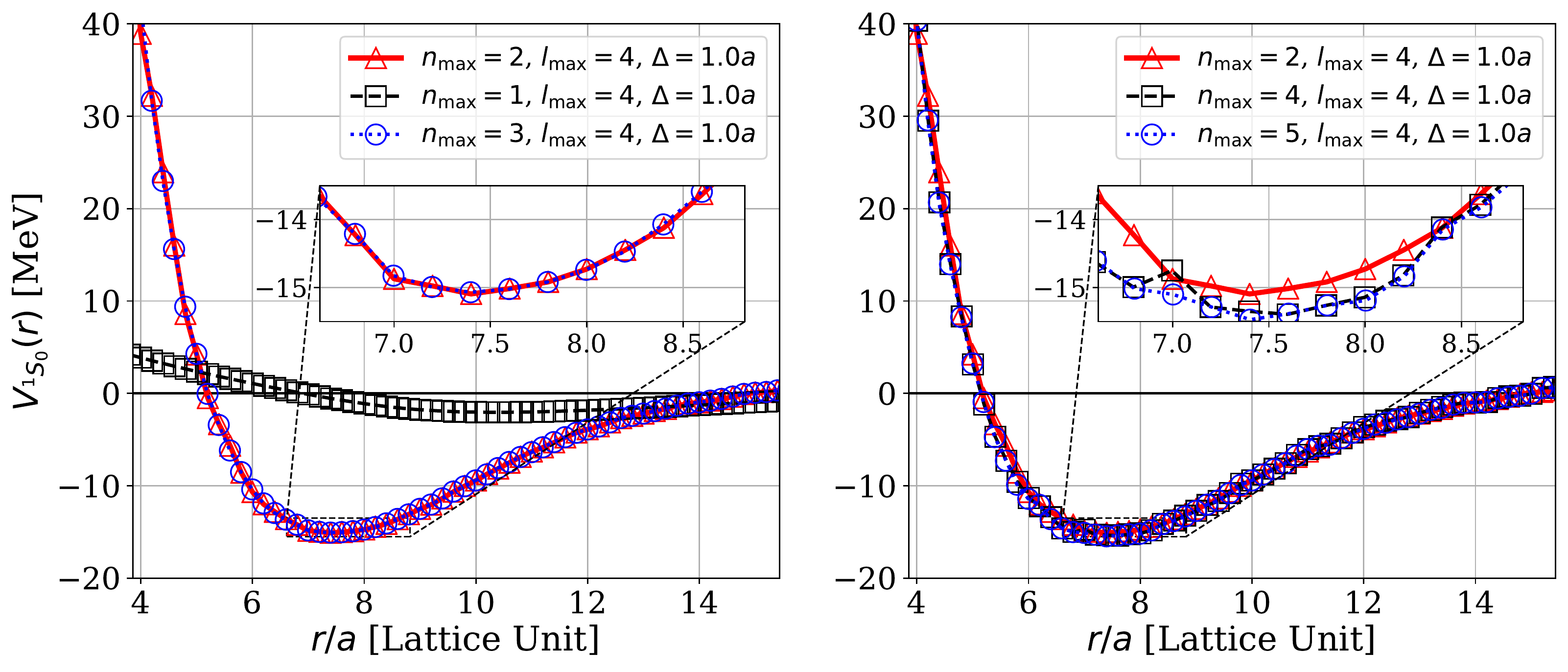} 
	\includegraphics[width=16cm]{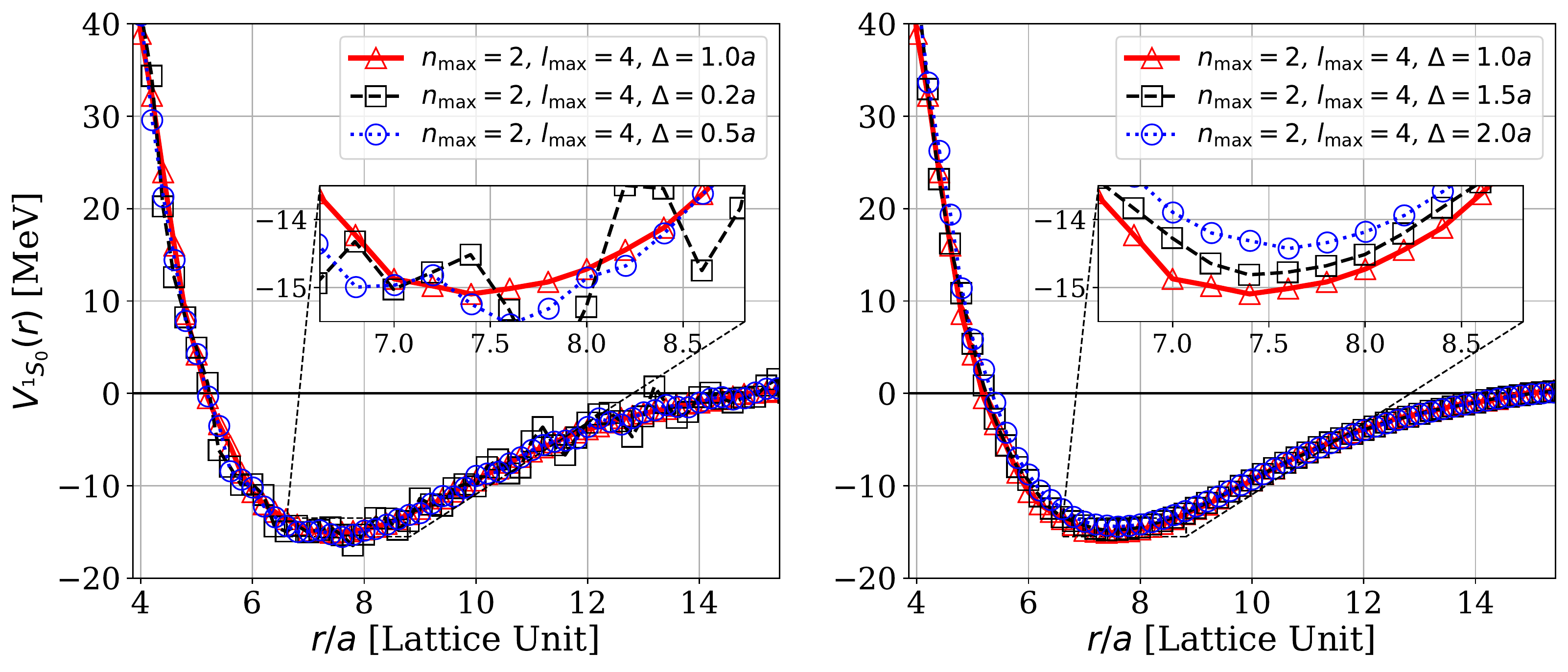} 
	\includegraphics[width=16cm]{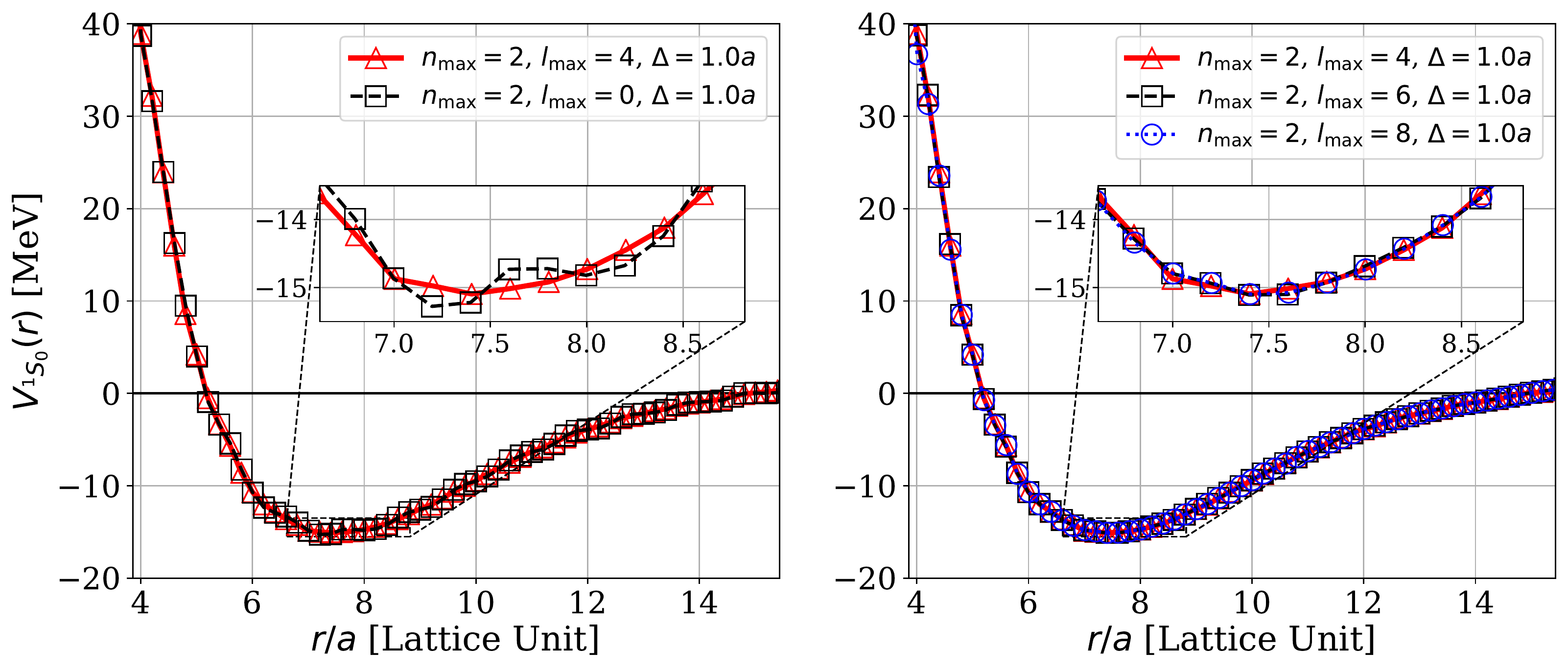} 
	\caption{The parameter dependence in Misner's method for the potential.
	  The red line (with triangles) shows the HAL QCD potential constructed with Misner's method with the parameters $n_{\mathrm{max}}=2$, $l_{\mathrm{max}}=4$, and $\Delta=a$, while the other lines (with other symbols) represent the one with parameters specified in the legends. \label{fig:param_dep}}
\end{figure}
We here discuss dependencies of potentials on various parameters in Misner's method,
$\Delta, n_{\mathrm{max}}$ and $l_{\mathrm{max}}$, which correspond to the thickness of the spherical shell, the maximum number of bases for the radial function and that for the spherical harmonics, respectively.
Throughout this section, the potential is constructed from the $l=0$ component of the NBS wave function.

We first show the $n_{\mathrm{max}}$ dependence with other parameters fixed to $\Delta=a$ and $l_{\mathrm{max}}=4$, in upper two figures of Fig.~\ref{fig:param_dep}, where we vary the value of $n_{\mathrm{max}}$ from $1$ to $5$. 
In Fig.~\ref{fig:param_dep}, we plot only the central values of the potentials without statistical errors in order to make it easier to see the dependence on parameters in Misner's method.
Note that the magnitude of statistical errors are found to be stable against changing parameters.
For $n_{\mathrm{max}} = 1$, the potential does not reproduce the correct behavior.
The reason is that the second derivative of the Legendre polynomial $P_n^{\prime\prime}(x)$ is zero for $n=0$ and $1$ so that the Laplacian term in the potential from the spherical harmonics amplitude vanishes.
The small contribution to the potential at $n_{\mathrm{max}} = 1$ comes from the time-derivative terms (the first and second terms in Eq.~\eqref{eq:1S0_pot_Rcorr}).
Thus it is necessary to take $n_{\mathrm{max}}\ge2$, for which we find that the potentials are almost stable against the change of $n_{\mathrm{max}}$.
While we observe small oscillations of the potential for $n_{\mathrm{max}} \ge 4$, they are probably due to the numerical instabilities associated with small eigenvalues of $\mathcal{G}_{AB}$ caused by a large number of $n_{\mathrm{max}}$, as discussed in Sec.~\ref{subsec:remarks}. Recall also
that the discretization errors of the radial orthonormal basis function $G^{R,\Delta}_n(r)$ is
known to be $\mathcal{O}(\Delta^{n_{\mathrm{max}}+2})$~\cite{Fiske:CQG2006}. 
Since the discretization errors in our lattice QCD action~\cite{Miyamoto:NPA2018} is $\mathcal{O}(a^2)$, a choice of the parameters $n_{\mathrm{max}}=2$ with $\Delta=\mathcal{O}(a)$ is reasonable. 

We next present the $\Delta$ dependence of the potential in middle two figures of Fig.~\ref{fig:param_dep}, where we take $\Delta=0.2a, 0.5a, a, 1.5a$ and $2a$ with $n_{\mathrm{max}}=2$ and $l_{\mathrm{max}}=4$ fixed.
For $\Delta=0.2a$ and $0.5a$, we find the comb-like structures even in the potential constructed from the $l=0$ component, while such structures are absent for the potential with $\Delta \ge a$.
This can be understood from the fact that the number of the data points in the spherical shell becomes too small for small $\Delta$ to reproduce the spherical harmonics amplitude accurately.
For larger $\Delta$, on the other hands, there appear small deviations from the potential with $\Delta =a$ at short distances, while the potentials at long distances are stable against the change of $\Delta$. 
This may be explained by the fact that variations of the $R$-correlator in the spherical shell become sizable for larger $\Delta$, so that we need to enlarge $n_{\mathrm{max}}$ accordingly to approximate the spherical harmonics amplitudes precisely.
In fact, by taking larger value of $n_{\mathrm{max}}$ in the case of $\Delta = 1.5a, 2a$, we find that the results tend to converge to that with $\Delta = a, n_{\mathrm{max}}=2$.
From these observations, we take $\Delta=a$ and $n_{\rm max}=2$ in this paper.

Finally, we investigate the $l_{\mathrm{max}}$ dependence with $\Delta$ and $n_{\mathrm{max}}$ fixed, as shown in lower two figures of Fig.~\ref{fig:param_dep}.
We take $l_{\mathrm{max}}=0,4,6$ and $8$, while keeping $\Delta=a$ and $n_{\mathrm{max}}=2$. 
The potentials are stable against the change of $l_{\mathrm{max}}$.
While the potential is rather reasonable even in the case of $l_{\mathrm{max}}=0$, small oscillations are observed in the potential for this case. 
Such oscillations are absent  for $l_{\mathrm{max}} \ge 4$, indicating that $l_{\mathrm{max}} = 4$ component has small but non-negligible contributions in the $R$-correlator, while $l_{\mathrm{max}} > 4$ components are sufficiently small.
The $l = 6$ component in the $R$-correlator obtained with $l_{\mathrm{max}} \ge 6$ is actually found to be negligible.
Therefore we take $l_{\mathrm{max}} =4$  in this paper, as a conservative choice to avoid numerical instabilities for larger $l_{\mathrm{max}}$.
\section{Phase shifts for the spin-singlet $\Lambda_cN$ system} \label{sec:observables}
\begin{figure}[htb] \centering
	\includegraphics[width=16cm]{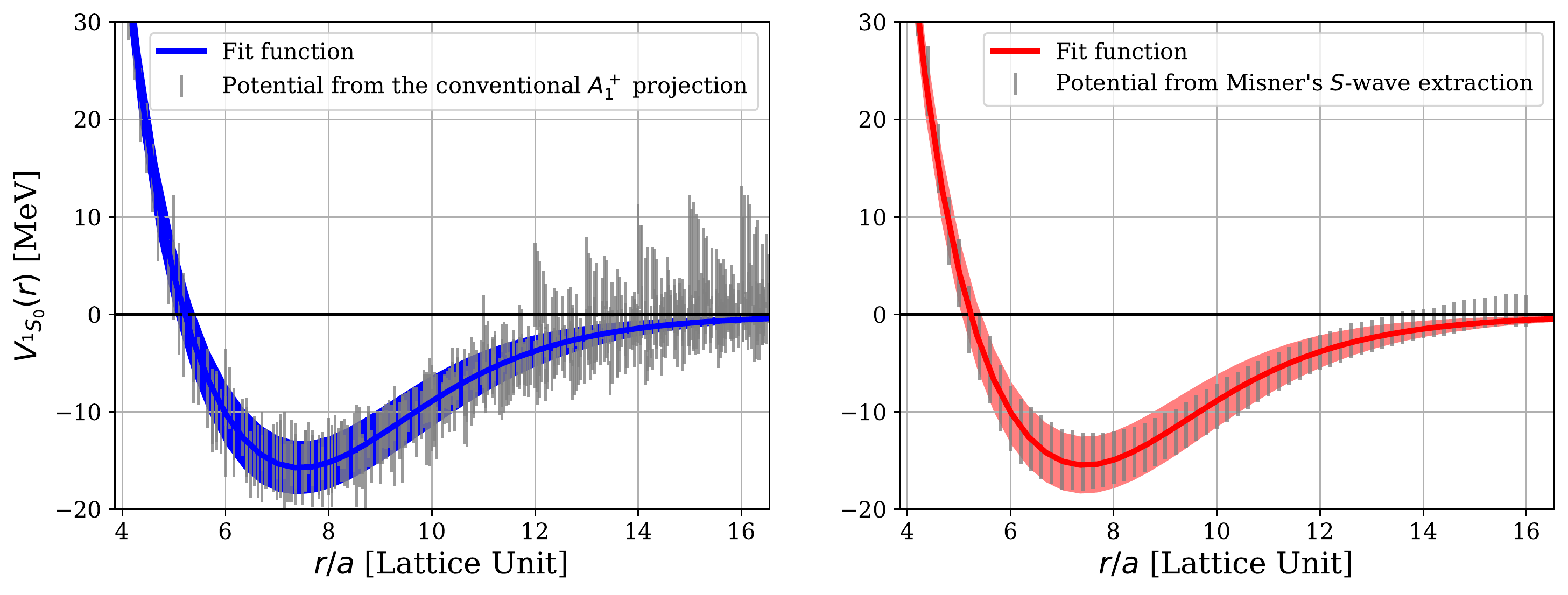} 
	\caption{Fits for the HAL QCD potential obtained
    from the conventional $A_1^+$ projection (left)  and 
    from Misner's $S$-wave extraction (right).
	  The fit-range is $r\in[2a,16a]$ for both cases. \label{fig:fitting_resutls}
  }
\end{figure}
We here compare the scattering phase shifts calculated from the HAL QCD potential
obtained by Misner's $S$-wave extraction 
with those
by the conventional $A_1^+$ projection, in order to estimate effects from $l\ge 4$ partial waves to physical observables.
For this purpose, we fit both potentials using the fit function
\begin{eqnarray}
	V_{\mathrm{fit}}(r) = a_1 e^{-\left(\frac{r}{a_2}\right)^2} + a_3 \left[ \left( 1-e^{-a_4 r^2} \right) \frac{e^{-a_5 r}}{r} \right]^2
	\label{eq:fit_func}
\end{eqnarray}
with the fitting range $r\in[2a,16a]$,
where both fit and original data are shown in Fig.~\ref{fig:fitting_resutls}.
While the conventional HAL QCD potential has large comb-like structures, the fit parameters $a_i$ are almost identical for both potentials, and, more surprisingly, the magnitude of statistical errors are also found to be similar. 
This observation indicates that $l \ge 4$ contributions in the conventional HAL QCD potential hardly affect the fit of the potential.
  The agreement for the fit parameters between two cases is most likely attributed to that
  the fit function for the potential is taken to be isotropy (See Eq.~(\ref{eq:fit_func})).
  In the fit, we employ the uncorrelated fit and a more systematic study with the correlated fit is left for future studies.

\begin{figure}[htb] \centering
	\includegraphics[width=12cm]{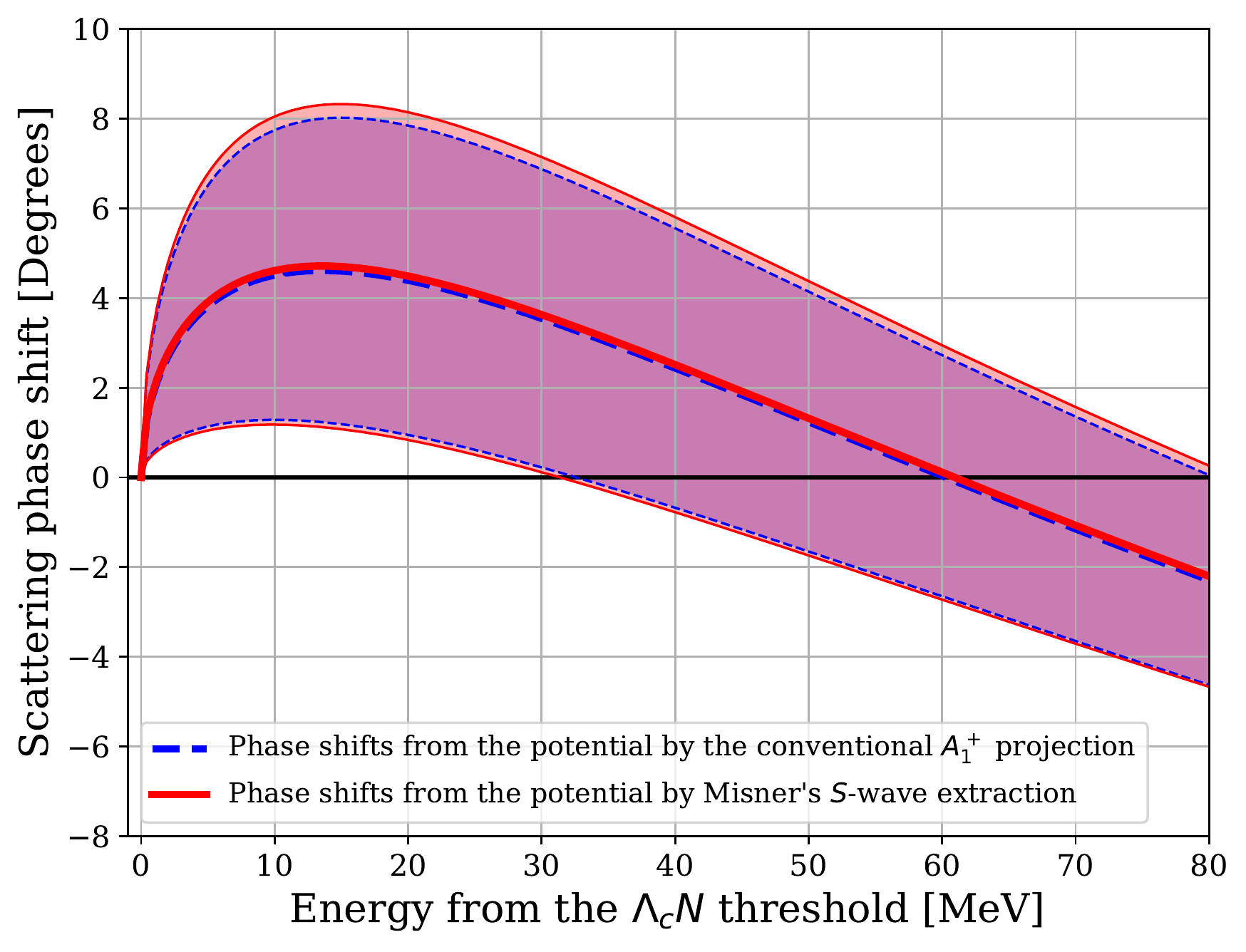} 
	\caption{Scattering phase shifts for the spin-singlet $\Lambda_c N$ system. 
	  The blue dashed line shows the phase shifts calculated
    from the potential by the conventional $A_1^+$ projection,
    while the red solid line represents the phase shifts calculated
    from the potential by Misner's $S$-wave extraction.
    \label{fig:phase_shift}
  }
\end{figure}

By solving the Schr\"odinger equation numerically with the fitted potentials, we extract the scattering phase shifts, which are shown in Fig.~\ref{fig:phase_shift}.
As expected from the fit results of the potentials, not only the central values of the scattering phase shifts but also their statistical errors are almost identical between two methods.

In the analysis with the conventional HAL QCD method, while the potential is affected by contaminations from higher partial waves with $l\ge 4$, it is confirmed that the results of the scattering phase shifts are not affected by such systematics for both central values and the magnitude of errors.
The conventional HAL QCD potential sometimes shows large  fluctuations, in particular for $NN$ channels at lighter pion masses.
The results in this paper tell us that these fluctuations are mainly systematic ones due to contaminations of higher partial waves enhanced by the second difference approximation of the Laplacian. 
By removing higher partial wave components from the $R$-correlator and calculating the Laplacian analytically,
the analysis with Misner's method reveals genuine statistical errors of the potential. 
An agreement in errors of  the scattering phase shifts (or equivalently the fitted potential) between
two analyses with the conventional method and Misner's method
provides a valuable check of the results.
\section{Summary and conclusion} 
\label{sec:summary}
In this paper, we have performed the approximated partial wave decomposition by Misner's method to lattice QCD data in order to extract the $l=0$ component of the $A_1^+$ projected NBS wave functions and its Laplacian for the $\Lambda_cN$ system in the spin-singlet channel, calculated in the $(2+1)$-flavor QCD on $(32a~\mathrm{fm})^3\times(64a~\mathrm{fm})$  at $m_\pi \simeq 700$ MeV~\cite{Miyamoto:NPA2018}.

We obtain the following results.
While the $A_1^+$ projected NBS wave functions contain small contaminations from $l\ge 4$ partial waves, 
such contaminations are enhanced if the Laplacian approximated by the second order difference
is applied to the NBS wave function,
which cause large fluctuations in the conventional HAL QCD potential.
With the use of Misner's method, since the Laplacian can be calculated analytically
for the designated ($l=0$) partial wave component in the NBS wave function, the potential is free from contaminations from higher partial waves, and thus its fluctuations become much smaller. 
Therefore, Misner's method is very useful to reduce superficial fluctuations of the potential.
If we fit the potentials as a function of $r$, not only the central values but also statistical errors of the fit parameters are almost  independent of whether the conventional HAL QCD potential or the potential extracted with Misner's method are used as input.
Consequently, the scattering phase shifts agree between two methods. 
This agreement demonstrates not only that Misner's method works well in the HAL QCD method but also the contaminations from higher partial waves in the study of $S$-wave scatterings are well under control even in the conventional HAL QCD method. 

Since one can approximately obtain the spherical harmonics amplitude for an arbitrary $l$ component by Misner's method, it is interesting to apply the method to extract the potentials from higher partial wave channels.
For example, in order to extract the tensor potential, one needs to obtain $l=0$ and $l=2$ components of the NBS wave function separately.
In the conventional HAL QCD method, one extracts the $l=2$ component from the NBS wave function by the projection, $(1-P^{A_1^+})$, which however also contains $l\ge4$ components. 
By employing Misner's method, on the other hand, one can extract $l=0$ and $l=2$ components separately without contaminations from higher partial waves, so as to obtain the tensor potential as well as the central potential without comb-like structures. 
\begin{center}
	--------------------------------------------
\end{center}
\section*{Acknowledgments}
Numerical data used in this study were obtained by the KEK supercomputer system (BG/Q) [Project number: 14/15-21, 15/16-12].
This work is supported in part by JSPS Grant-in-Aid for Scientific Research, No. JP19K03879, JP18H05236, JP18H05407, JP16H03978, JP15K17667, by a priority issue (Elucidation of the fundamental laws and evolution of the universe) to be tackled by using Post ``K" Computer, and by Joint Institute for Computational Fundamental Science (JICFuS).
The authors thank all the members of the HAL QCD Collaboration for discussion.
\appendix
\section*{Appendix}
\section{Approximated partial wave decomposition at fixed~$r$} \label{app:r-fix_Ldeco}
In this appendix, we propose a simpler method to extract $g_{00} (r)$ from the discrete data, which is compared with Misner's method.
We here consider the $A_1^+$ projected NBS wave function defined by
\begin{eqnarray}
	\psi^{A_1^+}(\vec{x}) &\equiv& P^{A_1^+} \psi(\vec{x}) \nonumber \\
	&=& Y_{00}^{A_1^+}(\theta,\phi) g_{00}(r) + \sum_{m=0,\pm4} Y_{4m}^{A_1^+}(\theta,\phi) g_{4m}(r) + \cdots, \label{eq:NBSwave_A1}
\end{eqnarray}
where $P^{A_1^+}$ is the projection operator to the $A_1^+$ representation, $Y_{00}^{A_1^+}(\theta,\phi)$ and $Y_{4m}^{A_1^+}(\theta,\phi)$ stand for the $A_1^+$ projected spherical harmonics, given by
\begin{eqnarray}
	Y_{00}^{A_1^+} (x,y,z) &=& Y_{00} (x,y,z) = \frac{1}{\sqrt{4\pi}} \\
	Y_{40}^{A_1^+} (x,y,z) &=& \frac{7}{8\sqrt{\pi}} \frac{x^4+y^4+z^4-3(x^2y^2+y^2z^2+z^2x^2)}{r^4} \\
	Y_{4,+4}^{A_1^+} (x,y,z) &=& Y_{4,-4}^{A_1^+} (x,y,z) = \sqrt{\frac{5}{14}} Y_{40}^{A_1^+} (x,y,z). \label{eq:sph_A1_def}
\end{eqnarray}
and the ellipsis denotes higher angular momentum components such as $l=6,8,\cdots$.
In the $A_1^+$ representation, the $l=4$ components become non-zero only for $m=0,\pm 4$.
Since $Y_{40}^{A_1^+}$ (also $Y_{4,\pm4}^{A_1^+}$) has an angular dependence, the NBS wave function at given $r$ is multi-valued, which make comb-like structures in the potential on the radial coordinate. 

Let us assume that there are $N$ points $\vec x_i$~$(i=1,\cdots,N)$ which satisfy $|\vec{x_i}|=r$ but can not be transformed each other by the cubic rotation. 
Neglecting components with $l\geq6$, we have
\begin{eqnarray}
	\psi^{A_1^+}(\vec{x_i}) &\simeq & Y_{00}^{A_1^+} g_{00}(r) + \sum_{m=0,\pm4} Y_{4m}^{A_1^+}(\vec{x}_i) g_{4m}(r) 
	= Y_{00}^{A_1^+} g_{00}(r) + Y_{40}^{A_1^+}(\vec{x}_i) g_{4}(r), \label{eq:NBSwave_A1_each_point}
\end{eqnarray}
where we omit the arguments for the constant $Y_{00}^{A_1^+}$, and $g_4(r) \equiv g_{40}(r) +\sqrt{\frac{5}{14}}(g_{44}(r) + g_{4-4}(r))$.
Eq.~\eqref{eq:NBSwave_A1_each_point} can be compactly written as
\begin{eqnarray}
	\left(
	\begin{array}{c}
		\psi^{A_1^+}(\vec{x}_1) \\
		\vdots \\
		\psi^{A_1^+}(\vec{x}_N)
	\end{array}
	\right) =
	\left(
	\begin{array}{cc}
		Y_{00}^{A_1^+} & Y_{40}^{A_1^+}(\vec{x}_1) \\
		\vdots & \vdots \\
		Y_{00}^{A_1^+} & Y_{40}^{A_1^+}(\vec{x}_N) 
	\end{array}
	\right)
	\left(
	\begin{array}{c}
		g_{00}(r) \\
		g_4(r)
	\end{array}
	\right), \label{eq:NBSwave_A1_matrix}
\end{eqnarray}
where the matrix in the right-hands side is an $N \times 2$ rectangular matrix in general.
If $N >  2$, we solve Eq.~\eqref{eq:NBSwave_A1_matrix} by using the Singular Value Decomposition (SVD) for the rectangular matrix.
The SVD for a $M\times N~(M>N)$ rectangular matrix $A$ is denoted as $A = U \Sigma V^\dagger$, where  $U,V$ are unitary matrices and $\Sigma$ is a diagonal matrix for the singular values.
Then the generalized inverse matrix is defined as $A^{-1} = V \Sigma^{-1} U^\dagger$, where $\Sigma$ and $\Sigma^{-1}$ are given by
\begin{eqnarray}
	\Sigma &\equiv& \left(
	\begin{array}{c}
		\mathrm{diag}(\sigma_1, \cdots, \sigma_N) \\
		\mbox{\boldmath $0$}_{(M-N)\times N}
	\end{array}
	\right) \label{eq:SVD_Sigma_def} \\
	\Sigma^{-1} &\equiv&
	\left(
	\begin{array}{cc}
		\mathrm{diag}(\sigma^{-1}_1, \cdots, \sigma^{-1}_N) &
		\mbox{\boldmath $0$}_{N\times (M-N)}
	\end{array}
	\right), \label{eq:SVD_Sigma_inv_def}
\end{eqnarray}
where $\mbox{\boldmath $0$}_{M\times N}$ represents the $M\times N$ zero matrix.
We here assume that all singular values are non-zero, otherwise the generalized inverse matrix cannot be defined.

An extension to higher angular momentum components than $l=4$ is straightforward. 
Including the $l=6$ component, for instance, we can extract the radial functions for $l=0,4,$ and $6$ by solving the equation
\begin{eqnarray}
	\left(
	\begin{array}{c}
		\psi^{A_1^+}(\vec{x}_1) \\
		\vdots \\
		\psi^{A_1^+}(\vec{x}_N)
	\end{array}
	\right) =
	\left(
	\begin{array}{ccc}
		Y_{00}^{A_1^+} & Y_{40}^{A_1^+}(\vec{x}_1) & Y_{60}^{A_1^+}(\vec{x}_1) \\
		\vdots & \vdots & \vdots \\
		Y_{00}^{A_1^+} & Y_{40}^{A_1^+}(\vec{x}_N) & Y_{60}^{A_1^+}(\vec{x}_N)
	\end{array}
	\right)
	\left(
	\begin{array}{c}
		g_{00}(r) \\
		g_4(r) \\
		g_6(r)
	\end{array}
	\right). \label{eq:NBSwave_A1_matrix_upto_L6}
\end{eqnarray}
Since the matrix in the right-hands side is a $N\times 3$ rectangular matrix, we need at least 3 points which satisfy $|\vec{x}_i| = r$ but are not transformed by the cubic rotation.

We next consider the extraction of the Laplacian for the radial function such as $g_{00}(r)$ in order to construct the potentials.
Neglecting components with $l\geq6$ again, the Laplacian of the NBS wave function in Eq.~\eqref{eq:NBSwave_A1_each_point} becomes 
\begin{eqnarray}
	\vec{\nabla}^2 \psi^{A_1^+}(\vec{x_i}) &=& Y_{00}^{A_1^+} \vec{\nabla}^2 g_{00}(r) + \vec{\nabla}^2 \left[ Y_{40}^{A_1^+}(\vec{x}_i) g_{4}(r) \right]. 
	\label{eq:Lap_NBSwave_A1_each_point}
\end{eqnarray}
where the second term is evaluated in the continuum relation as
\begin{eqnarray}
	\vec{\nabla}^2 \left[ Y_4^{A_1^+}(\vec{x}_i) g_{4}(r) \right] = Y_4^{A_1^+}(\vec{x}_i) \left[ \vec{\nabla}^2 - \frac{4(4+1)}{r^2} \right] g_{4}(r) 
	\label{eq:Laplacian_relation_continuum}.
\end{eqnarray}
Combining Eqs.~(\ref{eq:NBSwave_A1_each_point}) with (\ref{eq:Lap_NBSwave_A1_each_point}), we have
\begin{eqnarray}
	\left(
	\begin{array}{c}
		\psi^{A_1^+}(\vec{x}_1) \\
		\vec{\nabla}^2 \psi^{A_1^+}(\vec{x}_1) \\
		\vdots \\
		\psi^{A_1^+}(\vec{x}_N) \\
		\vec{\nabla}^2 \psi^{A_1^+}(\vec{x}_N)
	\end{array}
	\right) =
	\left(
	\begin{array}{cccc}
		Y_{00}^{A_1^+} & 0 & Y_{40}^{A_1^+}(\vec{x}_1) & 0 \\
		0 & Y_{00}^{A_1^+} & -\frac{4(4+1)}{r^2} & Y_{40}^{A_1^+}(\vec{x}_1) \\
		\vdots & \vdots & \vdots & \vdots \\
		Y_{00}^{A_1^+} & 0 & Y_{40}^{A_1^+}(\vec{x}_N) & 0 \\
		0 & Y_{00}^{A_1^+} & -\frac{4(4+1)}{r^2} & Y_{40}^{A_1^+}(\vec{x}_N) \\
	\end{array}
	\right)
	\left(
	\begin{array}{c}
		g_{00}(r) \\
		\vec{\nabla}^2 g_{00}(r) \\
		g_4(r) \\
		\vec{\nabla}^2 g_4(r)
	\end{array}
	\right), \label{eq:Lap_NBSwave_A1_matrix}
\end{eqnarray}
which can be solved by SVD.
Note that the Laplacian in the left-hand-side of Eq.~\eqref{eq:Lap_NBSwave_A1_each_point} is approximated by the second order difference, which has $\mathcal{O}(a^2)$ discretized errors.
\begin{figure}[bth] \centering
	\includegraphics[width=16cm]{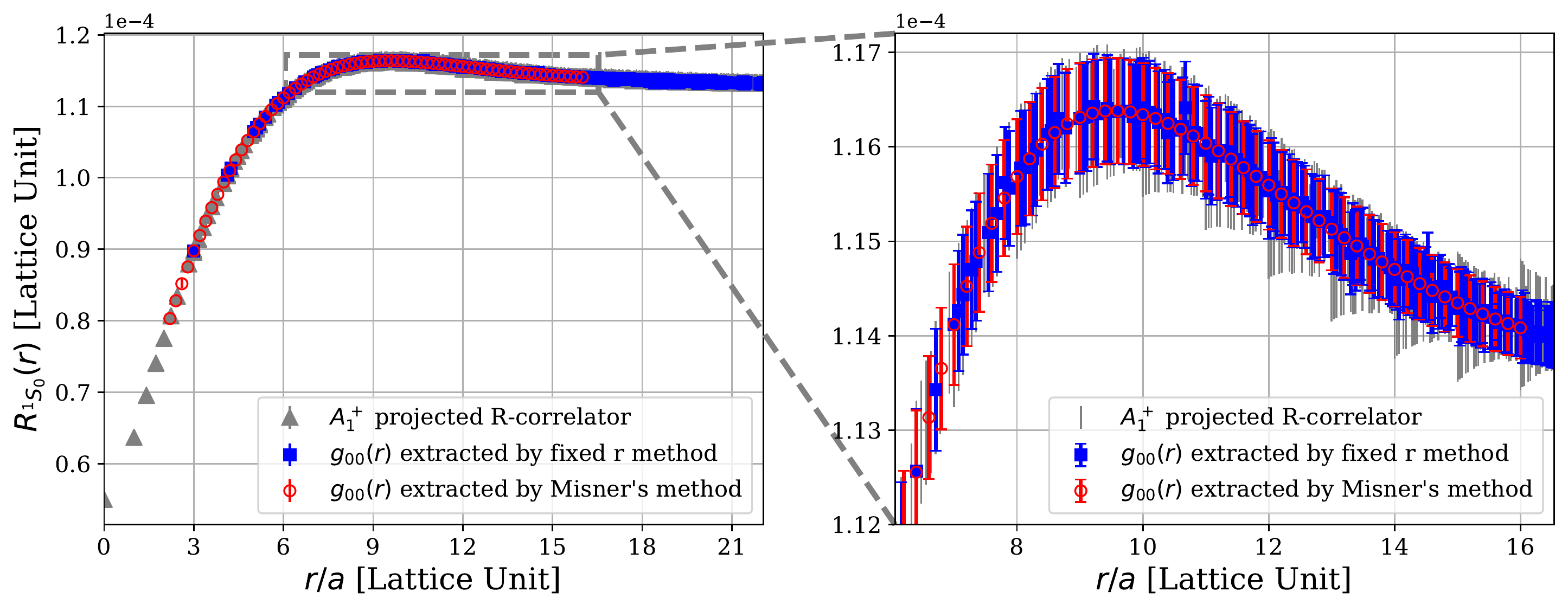}
	\includegraphics[width=16cm]{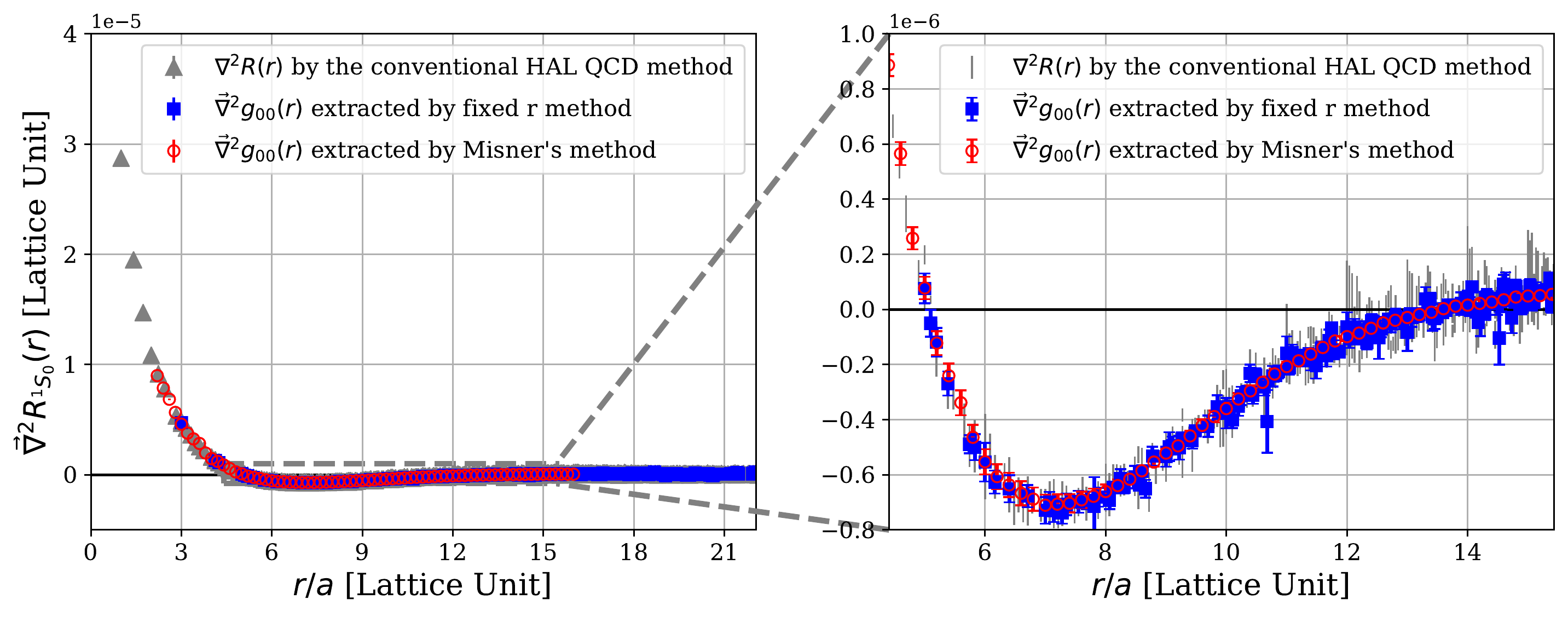} 
	\caption{The $R$-correlator  for the spin-singlet $\Lambda_c N$ system at $m_{\pi}\simeq $~700 MeV (Upper two figures) and its Laplacian term (Lower two figures).
	The $R$-correlator is calculated at $t-t_0 = 13a$. 
	The spherical harmonics amplitude $g_{00}(r)$ and its Laplacian term extracted by the method in this appendix are plotted (blue points), together with those in Misner's method (red points) 
	as well as  the original $A_1^+$ projected $R$-correlator and its Laplacian term (gray points).  
	 \label{fig:LcN_Rcor_1S0_t13_rfix}}
\end{figure}

\begin{figure}[htb] \centering
	\includegraphics[width=16cm]{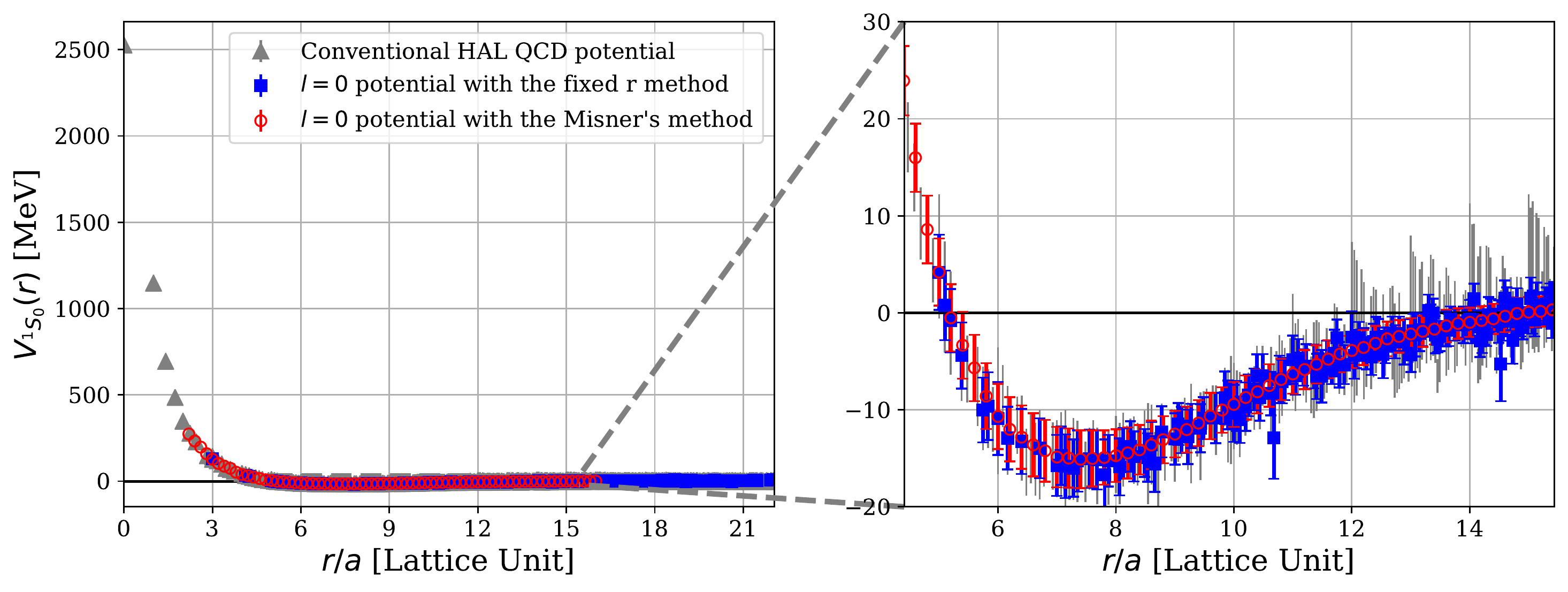} 
	\caption{The potential for spin-singlet $\Lambda_c N$ system. 
	The potential is constructed by the time-dependent HAL QCD method from the $A_1^+$ projected $R$-correlator calculated at $t-t_0 = 13a$ at $m_{\pi}\simeq $~700 MeV. 
The gray points show the potential calculated from the conventional HAL QCD method, while the blue and red points correspond to the potential constructed from the spherical harmonics amplitude $g_{00}(r)$ calculated by the method in this appendix and Misner's method, respectively. \label{fig:LcN_potential_1S0_t13_rfix}}
\end{figure}

An advantage of this method over Misner's method is that we need no parameter to extract the $l=0$ component of the NBS wave function. 
Applying this method, we extract the $l=0$ component of the $A_1^+$ projected $R$-correlator for the spin-singlet $\Lambda_cN$ system. 
In Fig.~\ref{fig:LcN_Rcor_1S0_t13_rfix}, $g_{00}(r)$ and its Laplacian are compared with those extracted by Misner's method.
The spherical harmonic amplitude $g_{00}(r)$ extracted by this method does not show comb-like structures and is consistent with $g_{00}(r)$ in Misner's method, while its Laplacian has comb-like structures, which probably originate from discretized errors due to the Laplacian in the left-hand-side of Eq.~\eqref{eq:Lap_NBSwave_A1_each_point}.
Consequently, the potential constructed from the $l=0$ component in this method, shown in Fig.~\ref{fig:LcN_potential_1S0_t13_rfix}, also has comb-like structures.
Therefore Misner's method works better for the HAL QCD potential than the method in this section, which however may be used to extract $g_{00}(r)$ only.\footnote{On this point, the method in this section essentially corresponds to Misner's method with the thickness of the spherical shell $\Delta \rightarrow 0$.
With the generalization of $\Delta \neq 0$, Misner's method can utilize more data points and thus obtain $g_{00}(r)$ at more points of $r$ (red) than the method in this section (blue) at short distances.}

\end{document}